\documentclass[preprint]{aastex62}
\usepackage{natbib}
\usepackage{graphicx}
\usepackage{epstopdf}
\usepackage{mathtools}
\usepackage{bm}
\newcommand{\be}{\begin{equation}}
\newcommand{\ee}{\end{equation}}
\newcommand{\ba}{\begin{eqnarray}}
\newcommand{\ea}{\end{eqnarray}}

\newcommand{\bvarepsilon}{\mbox{\boldmath$\varepsilon$}}
\newcommand{\nn}{\mbox{} \nonumber \\ \mbox{}}
\usepackage{longtable}

\begin{document}
\title{QED Phenomena in an Ultrastrong Magnetic Field. II. \\
Electron--Positron Scattering, $e^\pm$--Ion Scattering and Relativistic Bremsstrahlung}

\author{Alexander Kostenko}
\affiliation{Department of Astronomy and Astrophysics, University of Toronto, 50 St. George Street, Toronto, ON M5S 3H4, Canada}
\author{Christopher Thompson}
\affiliation{Canadian Institute for Theoretical Astrophysics, 60 St. George Street, Toronto, ON M5S 3H8, Canada}

\received{2018 September 7}
\published{2019 April 10; Erratum added}
\submitjournal{ApJ}
\shorttitle{QED in an Ultrastrong Magnetic Field II.}
\shortauthors{Kostenko \& Thompson}


\begin{abstract}
This paper continues the approach of \cite{KT2018} to calculating quantum electrodynamic processes
in the ultrastrong magnetic field near some neutron stars, such as magnetars or merging binary neutron stars.   Here we
consider electron-positron scattering, the Coulomb scattering of electrons and positrons off ions, and relativistic 
$e^\pm$-ion bremsstrahlung.   The evaluation of differential and total cross sections simplifies considerably
when the magnetic field lies in the range $10^3B_{\rm Q} \gg B \gg B_{\rm Q}$, where $B_{\rm Q} \equiv m^2/e = 
4.4\times 10^{13}$ G.  Then, relativistic motion of $e^\pm$ is possible even when restricted to the
lowest Landau state.   Accurate results for differential and total cross sections are obtained by truncating the sum over
intermediate-state Landau levels and otherwise disregarding terms inversely proportional to the magnetic field, which are
complicated enough to have inhibited previous attempts to calculate magnetic electron-positron scattering and relativistic
bremsstrahlung.  A quantitative account is made of the effects of Debye screening.
\end{abstract}


\section{Introduction}

An unusual electrodynamic regime is encountered near the surfaces of some neutron stars, e.g. magnetars, where free
electrons and positrons may be largely restricted to longitudinal motion along a very strong magnetic field.   
The relative rates of various electrodynamic processes can be altered dramatically by the presence of a magnetic field
exceeding $B_{\rm Q} = m^2/e = 4.4\times 10^{13}$ G, where $m$ and $-e$ are the electron mass and charge, respectively.  The
calculations presented here are partly motivated by a desire to
understand the bright nonthermal X-ray emission of magnetars, which extends with rising intensity to at least 100 keV
\citep{Kuiper2006,Mereghetti2006}, and the onset of pair fireballs around bursting magnetars \citep{TD2001} and 
inspiraling and colliding neutron stars \citep{HL2001}.

A companion paper \citep{KT2018} gives a simplified, but accurate, description of electron-photon scattering,
electron-positron pair creation, and pair annihilation in ultrastrong magnetic fields.  In the regime considered, 
incoming, outgoing, and internal electron lines are all restricted to the lowest Landau level.
Here a similar approach is adopted for Coulomb scattering of electrons and positrons off ions, relativistic 
$e^\pm$-ion bremsstrahlung, and electron-positron scattering.  The complications introduced by performing the full sum
over intermediate-state Landau levels appear to have inhibited previous attempts to calculate the last two processes.
We are able, for the most part, to present results for differential and total cross sections in terms of compact
analytic formulae.  
In each case, an ab initio calculation is performed in the regime where
relativistic electrons and positrons may be inhibited from Landau transitions ($B > B_{\rm Q}$) but nonlinearities
due to vacuum polarization are of secondary importance ($B \ll 10 \alpha_{\rm em}^{-1}\,B_{\rm Q} \sim 10^3\,B_{\rm Q}$,
where $\alpha_{\rm em} \simeq 1/137$ is the fine structure constant).  

The basic rates of magnetic electron-photon scattering and electron-positron annihilation (as reviewed by 
\citealt{HL2006, KT2018}) are well covered in the literature.   Nonetheless, our simplified approach permitted a
detailed examination of secondary effects, which have a strong influence on overall rates, including the conversion
of final-state photons to electron-positron pairs (which is kinematically possible in the presence of the magnetic
field;  \citealt{Erber1966, LLT4, DH1983, KT2018})  and the enhancement in scattering associated with a $u$-channel pole involving 
an intermediate-state positron. 

As for the processes considered here, electron-positron scattering in a strong magnetic field is treated for the
first time.  The Coulomb scattering of nonrelativistic electrons by fixed ions into a range of Landau states
was considered by \cite{ventura1973}, \cite{PL1976}, and \cite{Miller1987} and generalized to include relativistic electron motion
by \cite{Bussard1980} and \cite{Langer1981}.  Free-free emission by nonrelativistic electrons and static ions was calculated to all
orders in the background magnetic field by \cite{PP1976}, and a simplified treatment applicable to sub-QED 
magnetic fields was also given by \cite{Canuto1969} and \cite{Lieu1983}.  
Quantized Landau excitations of the ion, which may be followed by rapid radiative de-excitation, lead to important modifications
of Coulomb scattering and bremsstrahlung.  Ion recoil without Landau excitations was included by \cite{Neugebauer1996}.  
A full treatment of ion Landau excitations during both Coulomb scattering and free-free emission has been given, in the case
of nonrelativistic electrons, by \cite{PC2003}, \cite{PL2007}, and \cite{Potekhin2010}.

The introduction of super-QED magnetic fields now leads to a second simplification:
not only does a transrelativistic electron remain confined to its lowest Landau level, but, in addition, the
cross section for the transition of a proton up (or down) in Landau level is substantially reduced.   For both $e^\pm$-ion
scattering and free-free emission, this reduction is by a factor of $\sim B_{\rm Q}/B$.  This behavior has a heuristic
semiclassical explanation:  the excitation of a proton to the first Landau level requires that it gain
a gyrational momentum $p_\perp^2 = 2eB = 2(B/B_{\rm Q}) m^2$, but a proton receives at most momentum $\sim m$ by recoil
from a transrelativistic electron moving along the magnetic field.  This means that collisionally induced emission
at the proton cyclotron resonance (around 10 keV near the surface of a magnetar) need not overwhelm bremsstrahlung
at 30-300 keV.  

These two simplifications allow us to obtain a closed-form expression for the bremsstrahlung
cross section with relativistic electron motion and an immobile ion.  A calculation of relativistic free-free emission by
$e^\pm$ interacting with an ion already in an excited Landau state is beyond the scope of this paper.  Depending on 
the astrophysical context, either relativistic motion of electrons or Landau quantization of ions will be the
more important effect, in which case either the present results or those of \cite{PL2007} will be more relevant.

The electrostatic interaction between electrons and positrons (forming positronium atoms; \citealt{SU1985}) or 
between electrons and protons (forming hydrogen atoms;  \citealt{HL2006}, and references therein) is enhanced by
a strong magnetic field.  Nonetheless, it will have a small effect on the cross section for Coulomb scattering
of transrelativistic electrons and positrons.   We quantify here for the first time its effect on relativistic free-free
emission, where the correction is found to be at the $\sim 10\%$ level.

This paper is organized as follows.  Section \ref{s:qed} contains an abridged overview of our procedure for evaluating
matrix elements in a background magnetic field.  Each calculation is performed using the magnetized
electron and positron wave functions introduced by \cite{ST1968} and \cite{MP1983}.

Electron-positron scattering is considered in Section \ref{BhabhaSection}.  The annihilation channel ($s$-channel)
contains a pole that is modified by the decay of the intermediate-state photon into a pair.  By contrast, electron-electron scattering 
\citep{Langer1981} is uninteresting when all initial- and final-state particles are
confined to the lowest Landau state.

Next, in Section \ref{EpmIonSc}, we show how Debye screening modifies the cross section for
relativistic $e^\pm$ scattering off an immobile ion;  in a one-dimensional plasma, screening has an important buffering
effect on scattering at large impact parameters.  A comparison is also made with the classical treatment, where complicated
chaotic behavior is encountered in the case of electron-ion scattering \citep{Hu2002}.

Lastly, we consider relativistic bremsstrahlung in the Born approximation in Section \ref{s:bremss}.  
The Gaunt factor is evaluated over a wide range of frequencies, and an estimate is given of the error caused by the neglect
of the electron-ion interaction in the electron wave function (Section \ref{s:born}).  Free-free
emission is also compared with radiative recombination into a bound hydrogen atom.

We adopt natural units ($\hbar = c = k_B = 1$) throughout this paper, along with the $(+---)$ metric signature.  The
Dirac gamma matrix convention matches that used by \cite{MP1983},
\be
\gamma^0 = \begin{pmatrix}
1 & 0 \\ 
0 & -1 
\end{pmatrix};\quad\quad
\gamma^j = \begin{pmatrix}
0 & \sigma^j  \\ 
-\sigma^j  & 0 
\end{pmatrix}. 
\ee
Here, each $0$ and $1$ element denotes a $2\times 2$ matrix, and $\sigma^j$ are the usual Pauli matrices.  
Landau gauge ${\bm A} = Bx \hat y$ is chosen for the background vector potential, and we alternatively use
Cartesian coordinates $(x, y, z)$ and spherical coordinates $(\theta,\phi)$ (with the axis $\theta = 0$
aligned with $+\hat z$) to describe the wavevectors of interacting particles.

\section{Rules for Computing QED Processes in a Background Magnetic Field}\label{s:qed}

This section summarizes basic properties of electron, positron, and photon wave functions in a background magnetic field,
along with the coordinate space Feynman rules.  A more complete description can be found in 
\cite{KT2018}.

1. The photon wave function is written as
\be
A^\mu(x^\nu) = {\varepsilon^\mu\over (2\omega L^3)^{1/2}} e^{-ik\cdot x};  \quad\quad  k^\mu = \omega(1, \hat k),
\ee
where $k^\mu$ is the wave 4-vector and $L^3$ is the normalization volume.  

2. The two photon polarization states are labeled as the ordinary (O) mode,
\be\label{eq:pol}
\varepsilon_{\rm O}^z = \sin\theta; \quad \varepsilon_{\rm O}^\pm = \varepsilon_{\rm O}^x \pm i\varepsilon_{\rm O}^y = 
-\cos\theta e^{\pm i\phi},
\ee
with an electric vector partly aligned with the background magnetic field, $\bvarepsilon\cdot{\bm B} \neq 0$; 
and the extraordinary (E) mode,
\be
\varepsilon_{\rm E}^z = 0;  \quad \varepsilon_{\rm E}^\pm = \mp ie^{\pm i\phi},
\ee
with $\bvarepsilon\cdot{\bm B} = 0$ (see \citealt{MV1979, HL2006}).   When Landau resonances are kinematically forbidden, 
the polarization dependence of the processes we consider reduces to a dependence on $\varepsilon^z$, effectively decoupling the E-mode.

3. The electron and positron wave functions are written, following \cite{ST1968} and \cite{MP1983}, as
\be
\left [ {\psi}_\mp^{(\sigma)}(x^\mu) \right ]_{p_z,n,a} = 
\left\{\begin{matrix} 
e^{-ip\cdot x}\,u_{n,a}^{(\sigma)}({\bm x}) \quad\quad (\rm electrons);\\ 
e^{ip\cdot x}\,v_{n,a}^{(\sigma)}({\bm x}) \quad\quad (\rm positrons). \\ 
\end{matrix}\right. 
\ee
Here $\sigma = \pm 1$ labels the spin state; $a$, the $x$ coordinate of the center of gyration; and $p^\mu$, the momentum $4$-vector,
\be
p^\mu = (E, 0, p^y, p^z);  \quad\quad p_y = a qB = {\rm sgn}(q){a\over \lambda_B^2}\quad (q = \pm e),
\ee
where $\lambda_B \equiv (|e|B)^{-1/2} = (B/B_{\rm Q})^{-1/2} m^{-1}$.  Under charge conjugation, the sign of $p^\mu$ 
reverses, and so the gyration center remains fixed. The energy-momentum relation of an electron or positron state is
\be
E^2 = p_z^2 + m^2 + p_n^2;   \quad\quad   p_n^2 \equiv 2n|e|B \equiv E_{0n}^2 - m^2,
\ee
where $n \geq 0$ is the Landau level.  Introducing the harmonic oscillator wave functions $\phi_n$, 
\be
\phi_n(x-a) = \frac{1}{L(\pi^{\frac{1}{2}} \lambda_B 2^nn!)^{\frac{1}{2}}} H_n\left( \frac{x-a}{\lambda_B} \right)
e^{-(x-a)^2/2\lambda_B^2},
\ee
where $H_n$ is the order-$n$ Hermite polynomial, the positive-energy spinors $u_{n,a}^{(\sigma)}({\bm x})$ are
\be\label{eq:spinors}
u_{n,a}^{(-1)}({\bm x})=\frac{1}{f_n}\begin{bmatrix}
-ip_zp_n\phi_{n-1}\\
(E+E_{0n}) (E_{0n}+m)\phi_n
\\ -ip_n(E+E_{0n})\phi_{n-1}
\\ -p_z(E_{0n}+m)\phi_{n}
\end{bmatrix} ; \quad\quad
u_{n,a}^{(+1)}({\bm x})=\frac{1}{f_n}\begin{bmatrix}
(E+E_{0n}) (E_{0n}+m)\phi_{n-1}\\
-ip_z p_n\phi_n
\\ p_z(E_{0n}+m)\phi_{n-1}
\\ ip_n (E+E_{0n}) \phi_n
\end{bmatrix};
\ee
while the negative-energy spinors $v_{n,a}^{(\sigma)}({\bm x})$ are
\be
v_{n,a}^{(+1)}({\bm x})=\frac{1}{f_n}\begin{bmatrix}
-p_n (E+E_{0n})\phi_{n-1}\\
-ip_z(E_{0n}+m)\phi_n \\
-p_z p_n\phi_{n-1} \\
i(E+E_{0n}) (E_{0n}+m)\phi_n
\end{bmatrix} ; \quad\quad
v_{n,a}^{(-1)}({\bm x})=\frac{1}{f_n}\begin{bmatrix}
-ip_z(E_{0n}+m)\phi_{n-1} \\
-p_n (E+E_{0n}) \phi_n \\
-i(E+E_{0n}) (E_{0n}+m)\phi_{n-1} \\
p_z p_n\phi_{n} \\
\end{bmatrix}.
\ee
Here we have introduced $f_n = 2\sqrt{E E_{0n}(E_{0n} + m)(E_{0n} + E)}$.  As is shown by \cite{KT2018},
the spinors with finite $p_z$ are obtainable by a Lorentz boost from the state $p_z = 0$.

4. The overlap of the photon wave function (wavevector $k^\mu$) with a pair of harmonic oscillator 
wave functions yields a factor $e^{-\lambda_B^2(k_x^2 + k_y^2)/4} \equiv e^{-\lambda_B^2k_\perp^2/4}$:
\ba\label{eq:phiint}
&& \int d^3x e^{i{\bm k}\cdot{\bm x}}\phi_n (x-a) \phi_0(x-b)e^{-iay/\lambda_B^2}
\left (e^{-iby/\lambda_B^2} \right)^* e^{ipz}\left ( e^{iqz} \right )^* \nn
&&\quad\quad = \frac{(2\pi)^2}{(2^n n!)^\frac{1}{2}} \delta\left ( k_y - \frac{a-b}{\lambda_B^2} \right ) 
\delta(k_z +p -q)e^{-\lambda_B^2 k_\perp^2/4}
e^{ik_x \left ( b+ \lambda_B^2 k_y/2 \right )} \lambda_B^n (-k_y + ik_x)^n,\nn
\ea
as derived by \citet{DB1980}.

5. A vertex between photon and electron lines is written as the spacetime integral
\ba
  && -ie\int d^4x \left[ \bar{\psi}_-^{(\sigma_I)}(x) \right ]_{p_{z,I}, n_I, a_I} \gamma_\mu A^\mu(x) 
\left [\psi_-^{(\sigma_i)}(x) \right]_{p_{z,i}, n_i, a_i} \nn
        &&\quad\quad  = -{ie\over (2\omega L^3)^{1/2}} \int d^4x \,e^{-i(p_i \pm k - p_I)\cdot x} \,
      \bar u_{n_I,a_I}^{(\sigma_I)}({\bm x}) \gamma_\mu \varepsilon^\mu  u^{(\sigma_i)}_{n_i,a_i}({\bm x}).
\ea
Here $i$ and $I$ label incoming and internal positive-energy electron states, and the photon is either
absorbed (wavevector $+k^\mu$) or emitted ($-k^\mu$).  The vertex between an incoming electron
and an internal positron is obtained by substituting $-p_I$ and $\bar v^{(-\sigma_I)}_{n_I,a_I}$ for 
$p_I$ and $\bar u^{(\sigma_I)}_{n_I,a_I}$.  

6. An internal electron line is represented by the propagator in coordinate space, 
\ba\label{ComptProp}
G_F(x'-x) &=& -i\int L\frac{da_I}{2\pi \lambda_B^2}\int L\frac{dp_{z,I}}{2 \pi}
\sum_{n_I=0}^{\infty} \biggl[ \theta(t'-t)\sum_{\sigma_I} u_{n_I,a_I}^{(\sigma_I)}({\bm x}')
\bar{u}_{n_I,a_I}^{(\sigma_I)}({\bm x})e^{-iE_I(t'-t)}e^{i{\bm p}_I\cdot({\bm x}'-{\bm x})} \nn
&& \quad\quad - \theta(t-t')\sum_{\sigma_I} 
v_{n_I,a_I}^{(\sigma_I)}({\bm x}')\bar{v}_{n_I,a_I}^{(\sigma_I)}({\bm x})
e^{iE_I (t'-t)}e^{-i{\bm p}_I\cdot({\bm x}'-{\bm x})} \biggr].
\ea

7. An internal photon line is also represented by the propagator in coordinate space,
\ba
G_\gamma^{\mu \nu} (x'-x)  = i \eta^{\mu \nu} \int \frac{d^3q}{2 \omega (2 \pi)^3} 
\left [ \theta(t'-t) e^{-iq\cdot (x'-x)} + \theta (t-t') e^{iq\cdot (x'-x)} \right ],
\label{PhotProp}
\ea
where $\eta^{\mu\nu}$ is the metric tensor.

8. The combined integral over $t$ and $t'$ generates a combination of an energy delta function and an energy
denominator:
\ba
\mp i\int dt \int dt' \theta[\mp(t-t')] e^{i(E_f + \omega_f \mp E_I)t'}\, e^{-i(E_i \mp E_I)t}
= \frac{2\pi \delta(E_i - E_f - \omega_f)}{E_i \mp E_I}.
\ea
Here $i$ and $f$ label incoming and outgoing particles in the case of photon emission by an electron or positron.

9. The contraction of the polarization vector with $\gamma$ matrices is
\be\label{eq:amatrix}
\gamma_0\gamma_{\mu}\varepsilon_i^\mu = -\begin{pmatrix}
0 & 0 & \varepsilon_i^z  &\varepsilon_i^- \\ 
0 &0  & \varepsilon_i^+ & -\varepsilon_i^z \\ 
\varepsilon_i^z & \varepsilon_i^- & 0 &0 \\ 
\varepsilon_i^+ & -\varepsilon_i^z &0  &0 
\end{pmatrix};  \quad\quad  (i = {\rm O}, {\rm E}). 
\ee 


10. The matrix element $S_{fi}$ includes energy and momentum delta functions, which, once squared, are handled according to
(e.g. in the case of electron-positron scattering from initial state $i$ to final state $f$)
\ba
\left[2\pi \delta(p_{z-,i} + p_{z+,i} - p_{z-,f} - p_{z+,f})\right]^2 \;&\rightarrow&\;  
 L\cdot 2\pi \delta(p_{z-,i} + p_{z+,i} - p_{z-,f} - p_{z+,f}) ; \nn
\left[2\pi \delta\left(\frac{a_{+,i}-a_{-,i}-a_{+,f}+a_{-,f}}{\lambda_B^2}\right) \right]^2 \;&\rightarrow&\;  
L\cdot 2\pi \delta\left(\frac{a_{+,i}-a_{-,i}-a_{+,f}+a_{-,f}}{\lambda_B^2}\right) ; \nn
\left[2\pi \delta(E_{-,i}+E_{+,i} -E_{-,f}-E_{+,f})\right]^2 \;&\rightarrow&\;  
T\cdot 2\pi \delta(E_{-,i}+E_{+,i} -E_{-,f}-E_{+,f}).
\ea
Here $T$ is the normalization time. No delta function in $p_x$ appears for our choice of background gauge.  For the sake of
brevity, such a combination of delta functions will be written in the following way:
\be
\delta^{(3)}_{fi}(E,p_y,p_z).
\ee

11. Summing over the phase space of a final-state photon involves the integral
\be 
\int L^3 {\omega_f^2 d\omega_f d\Omega_f\over (2\pi)^3}, 
\ee
where $\Omega_f$ is the solid angle.  For a final-state electron or positron, there is no sum over the $x$-component of
momentum; hence, the integral
\be
{|e|B\over 2\pi} \int L da_f  \int L {dp_{z,f}\over 2\pi} = \int L{da_f\over 2\pi\lambda_B^2} \int L{dp_{z,f}\over 2\pi}.
\ee

\section{Electron-Positron scattering}\label{BhabhaSection}

\begin{figure}
\epsscale{0.8}
\plotone{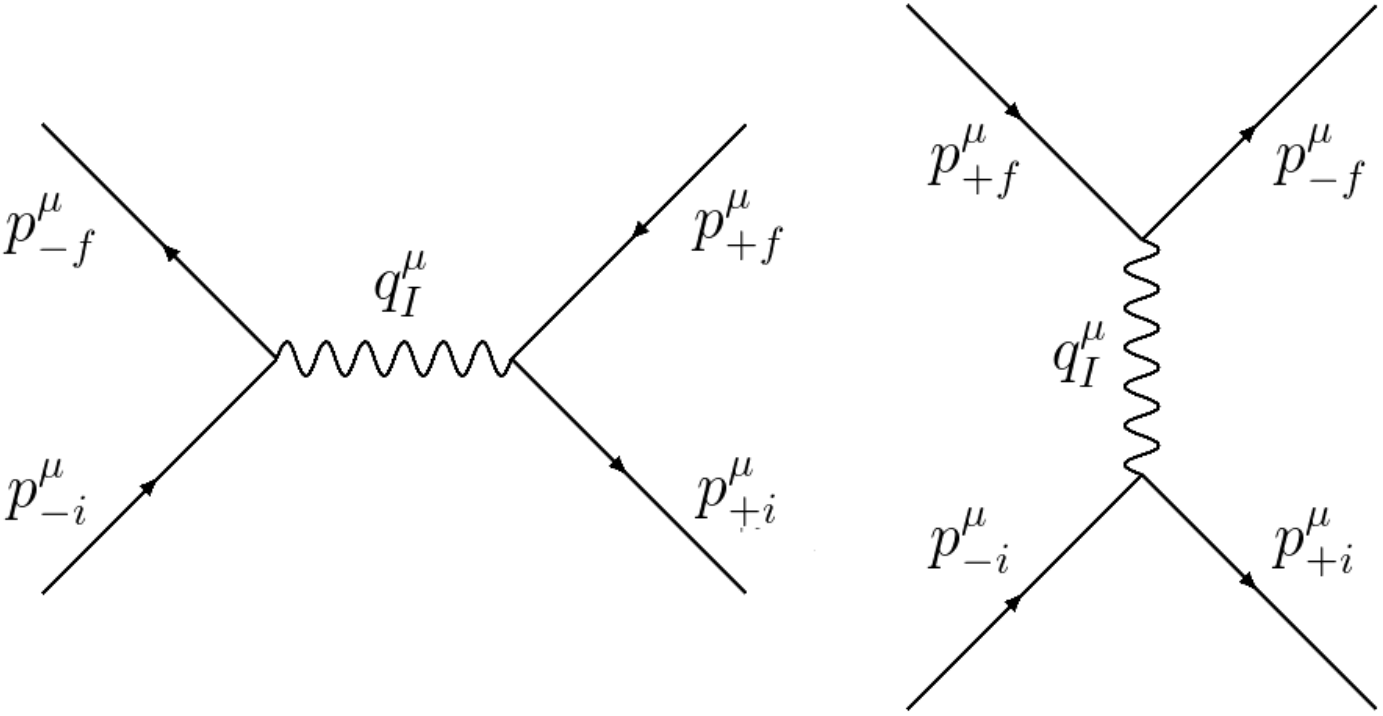}
\caption{Feynman diagrams for electron-positron scattering.  The left diagram [1] represents the
scattering ($t$) channel, and the right diagram [2] represents the annihilation ($s$) channel.}\label{BhabhaFeynman}
\end{figure}

Here we consider electron-positron (Bhabha) scattering in the presence of a magnetic field, 
\be
e^-_i + e^+_i \rightarrow  e^-_f + e^+_f,
\ee
with the initial and final particles all confined
to the lowest Landau state.   (Throughout this paper, the labeling of initial and final states is
summarized in the accompanying Feynman diagram;  see Figure \ref{BhabhaFeynman} for the case at hand.) The
particle kinetic energies are assumed to be well above the binding energy of positronium, which is \citep{SU2006} 
\be
|E_{\pm,0}| \simeq  \frac{\alpha_{\rm em}^2}{4} \left (\ln \frac{B}{B_{\rm Q}} \right )^2\, m
\ee
in a strong magnetic field.  Although the electron-positron interaction is important for adiabatic
conversion of a photon moving through a curved magnetic field into a pair \citep{SU1985}, the virtual photon appearing
in the scattering of warm electrons and positrons is far from the pair creation threshold.

Forward Coulomb scattering features a divergence \citep{Langer1981} that is resolved when 
one takes into account electric field screening \citep{PL2007}.  However, in the case where all particles are 
confined to the lowest Landau level, the conservation of momentum and energy implies that $p_{z-,f} = p_{z-,i}$, $p_{z+,f} = p_{z+,i}$.
The effects of forward scattering are therefore uninteresting, and we focus on backscattering of the electron 
and positron, corresponding to $p_{z-,f} = p_{z+,i}$, $p_{z+,f} = p_{z-,i}$.

The cross section is obtained from the integral
\be\label{eq:csbh}
\sigma = \frac{1}{|\beta_+-\beta_-|} \int \frac{da_{+,i} }{L} \int \frac{da_{-,f} L}{2 \pi \lambda_B^2}
\int\frac{da_{+,f} L}{ 2 \pi \lambda_B^2 } \int \frac{dp_{z-,f} L}{2 \pi}
\int \frac{dp_{z+,f} L}{2 \pi}\frac{L^3}{T}\Bigl|S_{fi}[1] + S_{fi}[2]\Bigr|^2,
\ee
where $\beta = p_z/E$.  The $S$-matrix has two terms, corresponding to the two diagrams in Figure \ref{BhabhaFeynman}. 
For the annihilation diagram, it is
\ba
S_{fi}[2] &=& -ie^2\int d^4x\int d^4x'\left [ \bar{\psi}_-^{(-1)}(x') \right ]_{p_{z-,f},n=0,a_{-,f}} 
\gamma_{\mu} \left [ \psi_+^{(+1)}(x') \right ]_{p_{z+,f},n=0,a_{+,f}} 
G_\gamma^{\mu \nu} (x'-x) \nn
&& \quad\quad \times \left [ \bar{\psi}_+^{(+1)}(x) \right ]_{p_{z+,i},n=0,a_{+,i}} 
\gamma_{\nu} \left [ \psi_-^{(-1)}(x) \right ]_{p_{z-,i},n=0,a_{-,i}}.
\ea
Substituting for the photon propagator from Equation (\ref{PhotProp}), this becomes
\be \label{BhabhaMEInt}
S_{fi}[2] =  \frac{ie^2}{(2 \pi)^2} \delta(E_{-,i}+E_{+,i} -E_{-,f}-E_{+,f})  
\int d^3q  \frac{I_{2,\mu} \eta^{\mu \nu}I_{1,\nu}}{(E_{+,i}+E_{-,i})^2-\omega^2},
\ee
where 
\be \label{I1Bhabha}
I_{1,\nu} = \int d^3x \left [ v_{0,a_{+,i}}^{(+1)*} ({\bm x})\right ]^T\gamma_0\gamma_{\nu}u_{0,a_{-,i}}^{(-1)}({\bm x})
e^{i({\bm p}_{+,i}+{\bm p}_{-,i})\cdot{\bm x}_\perp + i(p_{z+,i}+p_{z-,i})z} e^{-i{\bm q}\cdot{\bm x}},
\ee
and
\be \label{I2Bhabha}
I_{2,\mu} = \int d^3x' \left [ u_{0,a_{-,f}}^{(-1)*} ({\bm x}')\right ]^T\gamma_0\gamma_{\mu}v_{0,a_{+,f}}^{(+1)}({\bm x}')
e^{-i({\bm p}_{+,f}+{\bm p}_{-,f})\cdot{\bm x}'_\perp - i(p_{z+,f}+p_{z-,f})z'} e^{i{\bm q}\cdot{\bm x}'}.
\ee
These integrals are evaluated in Appendix \ref{BhabhaAppendix}.  

We work in the center-of-momentum frame, where $p_{z-,i} = -p_{z+,i} = p_z$, $\beta_{-,i} = -\beta_{+,i} = \beta$,
and $E_{-,i} = E_{+,i} = E$, giving
\be\label{eq:sfi1Int}
S_{fi}[2] = - i \frac{e^2m^2}{L^4 E^2} 
\int dq_x  \frac{\exp\left [ iq_x (a_{+,f} + a_{-,f} - a_{+,i} - a_{-,i})/2 - \lambda_B^2(q_x^2 + q_y^2)/2 \right ]}
    {4E^2- q_x^2 - q_y^2}\,(2\pi)^2\delta^{(3)}_{fi}(E,p_y,p_z).
\ee
Here the conservation of momentum implies $q_y  = q_{y,2} \equiv (a_{+,i} - a_{-,i})/\lambda_B^2$ and 
$q_z = p_{z+,i} + p_{z-,i} = 0$.

The matrix element for the scattering diagram can be obtained from $S_{fi}[2]$ by exchanging the momenta
of the initial-state positron and the final-state electron,  $p_{+,i}^\mu \leftrightarrow - p_{-,f}^\mu$ and 
$a_{+,i} \leftrightarrow a_{-,f}$, 
\be
S_{fi}[1] = -i\frac{e^2m^2}{L^4E^2}\, \int dq_x \frac{\exp\left [ iq_x (a_{+,f}+a_{+,i}-a_{-,f}-a_{-,i})/2 - 
\lambda_B^2(q_x^2 + q_y^2)/2 \right ]}{q_z^2+ q_y^2 + q_x^2} \, (2\pi)^2\delta^{(3)}_{fi}(E,p_y,p_z).
\ee
Here $q_z = p_{z-,i} - p_{z-,f} = 2p_z$ (for backscattering) and $q_y = q_{y,1} \equiv (a_{-,f} - a_{-,i})/\lambda_B^2$. 

To evaluate these integrals over $q_x$, we invoke the translational invariance of the background to set $a_{-,i} = 0$
and also take $\lambda_B \rightarrow 0$ in the exponentials.
We consider first the scattering diagram, which has poles at $ q_x = \pm i  \sqrt{4p_z^2 + q_{y,1}^2} $, so that the exponent
becomes $\exp[iq_x a_{+,i}]$.  Depending on the sign of $a_{+,i}$, we close the contour at positive or negative imaginary $q_x$, yielding
\be\label{eq:sfi2}
S_{fi}[1] = -i \frac{e^2m^2}{2L^4E^2} \exp\left [ - |a_{+,i}|  \sqrt{4p_z^2 + q_{y,1}^2}  \right ] 
\frac{{\rm sgn}(a_{+,i})}{\sqrt{4p_z^2 + q_{y,1}^2}}\,(2\pi)^3\delta^{(3)}_{fi}(E,p_y,p_z).
\ee

As for the annihilation diagram, we must take into account the finite width of the intermediate-state photon,
which has an energy exceeding $2m$.  The decay rate of a photon into a pair confined to the lowest Landau level is
\citep{KT2018}
\be\label{eq:31}
\Gamma_\pm(\omega,\theta) 
= 2\alpha_{\rm em} {B\over B_{\rm Q}}  {m^4\over \omega_\perp^2(\omega_\perp^2-4m^2)^{1/2}}
 e^{-(B_{\rm Q}/2B)(\omega_\perp/m)^2} \sin\theta; \quad\quad \omega_\perp = \omega\sin\theta.
\ee
Substituting $\omega = |{\bm q}| \rightarrow |{\bm q}| -  i\Gamma_\pm/2$, the integral in Equation (\ref{eq:sfi1Int}) 
gives
\be\label{eq:sfi1}
S_{fi}[2] =  \frac{e^2m^2}{2L^4E^2}\,\exp\Bigl[|a_{-,f}| (i q_{x,{\cal R}}^{\rm res} - q_{x,{\cal I}}^{\rm res})\Bigr] 
\frac{{\rm sgn}(a_{-,f})}{q_x^{\rm res}}\,(2\pi)^3\delta^{(3)}_{fi}(E,p_y,p_z).
\ee
Here $q_x^{\rm res} = q_{x,{\cal R}}^{\rm res} + iq_{x,{\cal I}}^{\rm res}$, where
\be
(q_{x,{\cal R}}^{\rm res})^2 = {1\over 2}(4E^2 - q_{y,2}^2) + {1\over 2}\sqrt{(4E^2 - q_{y,2}^2)^2 + \omega^2 \Gamma_\pm^2}; 
\quad\quad\quad 
(q_{x,{\cal I}}^{\rm res})^2 = -{1\over 2}(4E^2 - q_{y,2}^2) + {1\over 2}\sqrt{(4E^2 - q_{y,2}^2)^2 + \omega^2 \Gamma_\pm^2}.
\ee
Near the pole, one has $\omega = 2E$ to lowest order in $\Gamma_\pm$.

Substituting Equations (\ref{eq:sfi2}) and (\ref{eq:sfi1}) into Equation (\ref{eq:csbh}) gives the total cross section
\ba
\sigma &=& \frac{e^4m^4}{16\pi |p_z|E^3} \int da_{+,i} {da_{-,f}\over\lambda_B^2} {da_{+,f}\over\lambda_B^2}
 \int dp_{z-,f} dp_{z+,f} \,\delta^{(3)}_{fi}(E,p_y,p_z) \nn
&& \quad\quad \times \left | \frac{{\rm sgn}(a_{-,f})}{q_x^{\rm res}} 
\exp\Bigl[|a_{-,f}|(iq_{x,{\cal R}}^{\rm res} - q_{x,{\cal I}}^{\rm res})\Bigr] - i
\frac{{\rm sgn}(a_{+,i})}{\sqrt{4p_z^2 + q_{y,1}^2}} \exp\left [- |a_{+,i}|\sqrt{4p_z^2 + q_{y,1}^2} \right ]  \right |^2.\nn
\ea
The integrand involves four terms -- an annihilation term, a scattering term, and two cross terms. The cross terms, 
when integrated over $a_{+,f}$ using the delta function, yield terms of the form 
${\rm sgn}(a_{+,i}) {\rm sgn}(a_{-,f}) f(|a_{+,i}|,|a_{-,f}|)$, and vanish after further integration over 
$a_{+,i}$, $a_{-,f}$.  The remaining terms can be integrated in a straightforward manner, giving
\be \label{BhabhaSigmaFinal}
\sigma = \frac{\pi\alpha_{\rm em}^2 m^4}{4E^2 p_z^4} \left [ 1 + \beta^2 \frac{2E}{\Gamma_\pm}
\left ( \frac{\pi}{2} + \arcsin\frac{1}{\sqrt{1+\Gamma_\pm^2/4 E^2}} \right ) \right ].
\ee
This result can be self-consistently applied when the energy of the incoming electron and positron is too small
to permit excitation to the first Landau level, e.g. $p_z^2 < 2|e|B = 2(B/B_{\rm Q})m^2$.  
The $t$-channel dominates at low energy, with a Rutherford-like scaling in momentum,
$\sigma \propto p_z^{-4}$.  The $s$-channel begins to dominate when $\beta^2 \gtrsim \Gamma_\pm/m$.

\section{Scattering of Electrons and Positrons off Heavy Ions} \label{EpmIonSc}

\begin{figure}
\epsscale{0.35}
\plotone{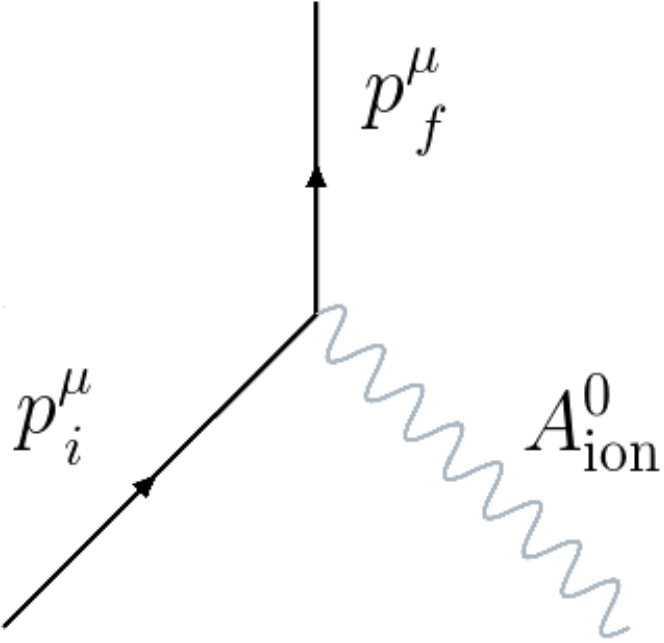}
\caption{Feynman diagram for electron-ion scattering. The ion is approximated as infinitely heavy and is
represented by a fixed Coulomb field (gray line).}\label{EPIon}
\end{figure}

Here we consider the scattering of relativistic electrons and positrons by heavy ions (Figure \ref{EPIon}), 
\be
Z+e^\pm_i \rightarrow Z+e^\pm_f,
\ee
in a magnetic field $B \gg B_{\rm Q}$.  The ion is treated as a fixed Coulomb field centered at $x = y = z = 0$,
\be
A^0_{\rm ion}({\bm x}) = \frac{Ze}{4 \pi \sqrt{x^2 + y^2 + z^2}}.
\ee
We assume that the ion has no gyrational motion in the initial state.  Excitation to a higher Landau level
during scattering is suppressed by a factor of $\sim B_{\rm Q}/B$ (compare Equations (8) and (9) of \cite{PL2007}
evaluated for final Landau state $N' = N+1 = 1$ as compared with $N' = N = 0$).

\subsection{Quantum Scattering}

We work in the Born approximation, where the deformation of the incoming and outgoing spinor wave functions by the 
Coulomb field is ignored.  In the case of electron-proton scattering, this is a good first approximation if the
initial kinetic energy of the electron is much larger than the binding energy of hydrogen in the strong magnetic field, 
which is \citep {HL2006}
\begin {equation} \label{MagHBind}
E_i - m \;\gg\; \left |E_{\rm H,0}  \right | \;\approx\; 0.32 \left ( \ln \frac{B}{\alpha_{\rm em}^2 B_{\rm Q}} \right )^2 
{\rm Ryd} \;=\; 8.7 \times 10^{-6}\left ( \ln \frac{B}{\alpha_{\rm em}^2 B_{\rm Q}} \right )^2 m.
\end {equation}

A characteristic impact parameter for the incoming particle is $a_i \sim 1/p_{z,i}$, which greatly
exceeds the localization length $\lambda_B$ of the electron or positron in the dimensions transverse to ${\bm B}$. As a result, given our
 choice of background magnetic gauge, the scattered particle can be assigned a well-defined impact parameter $x = a_i$ with respect to
 the  Coulomb center.  Focusing on electron-ion scattering, the matrix element written in coordinate space is
\be
S_{fi} = -ie\int d^4x\left [ \bar{\psi}_-^{(-1)}(x) \right ]_{p_{z,f},n=0,a_f} \gamma_{0} 
A_{\rm ion}^0({\bm x}) \left [\psi_-^{(-1)}(x) \right ]_{p_{z,i},n=0,a_i}.
\end {equation}
The spatial portion of this integral is
\ba \label{CoulombSpatialIntegral}
&& \int d^3x \left [ u_{0,a_f}^{(-1)*} ({\bm x}) \right ]^T A^0_{\rm ion}({\bm x})\, u_{0,a_i}^{(-1)}({\bm x}) \,
   e^{i ( {\bm p}_i - {\bm p}_f)\cdot{\bm x}_\perp + i(p_{z,i} - p_{z,f})z}\nn
   &=& {Ze\over 2} K_0[(p_{z,i} - p_{z,f})b] \, \frac{p_{z,i}p_{z,f} + (E_i + m)(E_f + m)}
       {L^2[E_i E_f (E_i + m)(E_f + m)]^{1/2}}
\, \delta\left ( \frac{a_f - a_i}{\lambda_B^2} \right ).
\ea
As we explain in Appendix \ref{BremssInteg}, the delta function in $a_i$ arises in the regime $a_i \gg \lambda_B$, 
meaning that over most impact parameters the scattered particle is only weakly deflected across the very strong 
magnetic field.   

We therefore focus on the case of backscattering, $p_{z,f} = -p_{z,i}$.  Performing the time integral gives
\begin {equation}
S_{fi} = -\frac{iZe^2}{L^2} \frac{2 \pi m}{E_i} K_0(2p_{z,i}a_{i}) \, \delta\left(\frac{a_f-a_i}{\lambda_B^2}\right) 
\,\delta(E_i - E_f).
\end {equation}
The absence of a delta function in $p_z$ follows from our assumption of an immobile ion.  The cross section per unit area is
\begin {equation}
{1\over 2\pi a_i}\frac{d \sigma}{da_i} = \frac{1}{|\beta_{i}|} \frac{L}{T}\int L \frac{da_{f}}{2 \pi \lambda_B^2} 
\int_{-\infty}^{0}L \frac{dp_{z,f}}{2 \pi} \left | S_{fi} \right |^2 =  \frac{Z^2e^4}{4 \pi^2 \beta_{i}^2 \gamma_i^2}K_0^2(2p_{z,i} a_{i}),
\end {equation}
where $\gamma_i = E_i/m$ and $\beta_i = p_{z,i}/E_i$. 
The integral over impact parameter converges, giving a total cross section
\begin {equation}\label{ElIonSigma0}
\sigma = \frac{\pi Z^2 r_e^2}{\beta_{i}^4 \gamma_{i}^4}
\end {equation}
for backscattering of an electron off a heavy positive ion.  Here $r_e = \alpha_{\rm em}/m$ is the classical electron radius.
The cross section for positron-ion scattering, evaluated in the same Born approximation, is identical to Equation (\ref{ElIonSigma0}).

The dependence of $\sigma$ on the incoming particle momentum is as expected for relativistic Rutherford backscattering. 
It also lines up with the $B \rightarrow \infty$ limit of the electron-ion backscattering cross section derived in \cite{Bussard1980}.   

The effect of Debye screening on Coulomb scattering has previously been computed in the nonrelativistic regime by \cite{PL1976},
\cite{Neugebauer1996}, and \cite{PL2007}.   
It is straightforward to include in the present situation, where relativistic electrons or
positrons are confined to the lowest Landau level.  Allowing for a background electron gas of number density $n_e$ and 
temperature $T$, the ion's electric field is reduced by a factor $\exp(-r/r_D)$, where $r_D = (T/4 \pi n_e Z^2e^2)^{1/2}$.
Then, we find
\begin {equation} \label{ElIonSigma}
\sigma = \frac{\pi Z^2 r_e^2}{\beta_{i}^2\gamma_{i}^2} \frac{1}{\beta_{i}^2 \gamma_{i}^2 + 1/4m^2 r_D^2}.
\end {equation}
Taking the nonrelativistic limit, this reproduces the results of \cite{PL1976} and \cite{Neugebauer1996}.  The
softening of the momentum dependence of the cross section has an important effect on ion-electron drag when the electrons
have a one-dimensional momentum distribution.

\subsection{Classical Scattering}

\begin{figure}
\epsscale{0.85}
\plotone{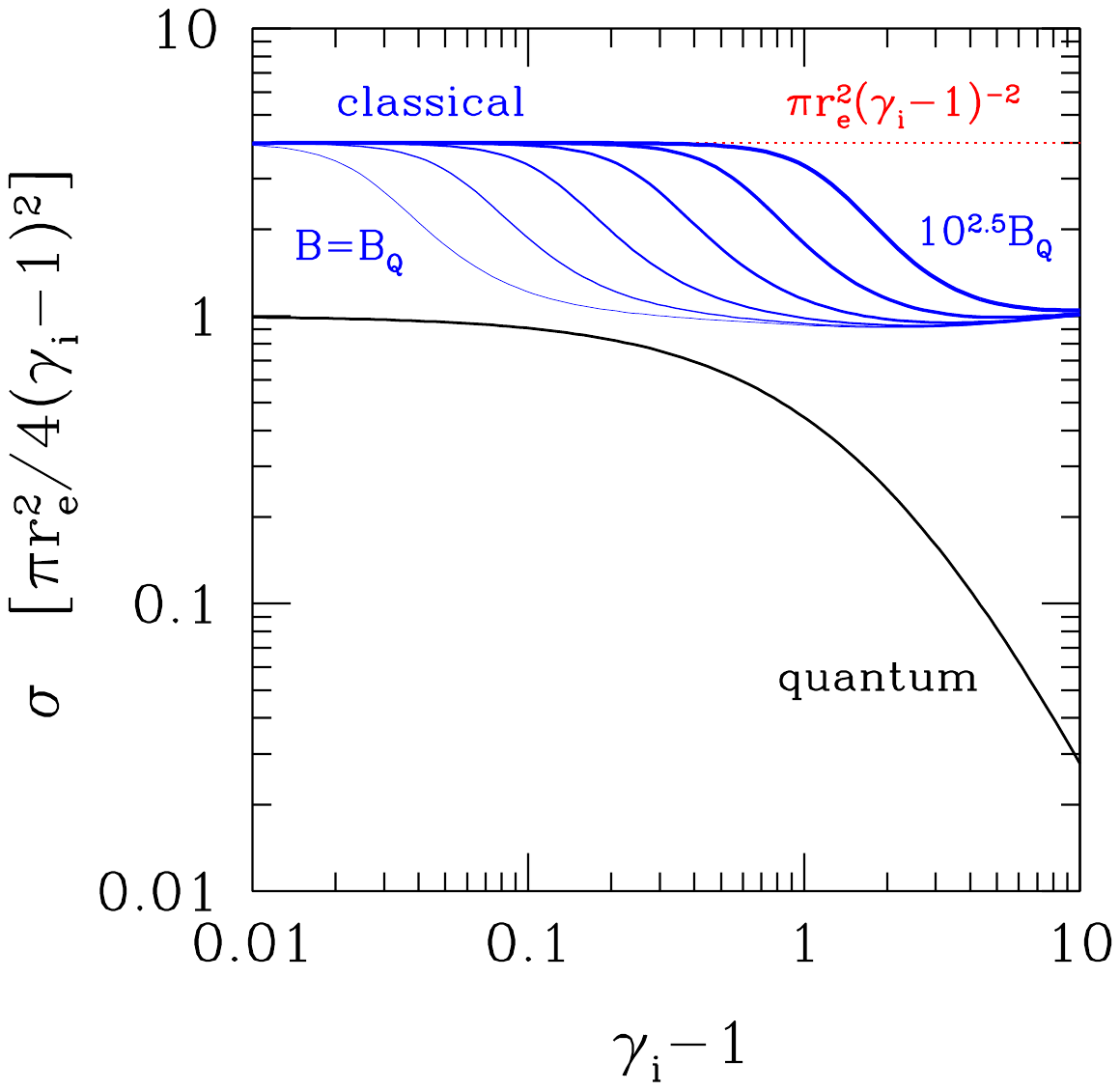}
\caption{Classical and quantum cross sections for the backscattering of a positron off a proton
in a strong magnetic field, vs. initial kinetic energy.  The positron is confined to the lowest Landau state,
and the proton is approximated as a fixed Coulomb potential.  As $B$ rises, the classical
cross section approaches the simple kinematic result $\pi r_e^2/(\gamma_{i} - 1)^2$; whereas it approaches the low-energy
quantum result when $\gamma_i-1 \gtrsim 0.1(B/B_{\rm Q})^{2/3}$.}\label{PositronIonClassical}
\end{figure}

There are dramatic differences in the classical backscattering of electrons and positrons off positive ions that are not evident
in the quantum Born calculation.   The result for positron-proton scattering (like charges) is shown in Figure
\ref{PositronIonClassical}.   As the background magnetic field $B \rightarrow \infty$, the cross section approaches the
simple kinematic result $\pi r_e^2/(\gamma_{i} - 1)^2$, which is obtained by treating the light charge
as a bead on a wire and finding the critical impact parameter where the Coulomb repulsion absorbs its asymptotic
kinetic energy.  On the other hand, the classical cross section approaches the low-energy quantum result when
$\gamma_i - 1 \gtrsim 0.1 (B/B_{\rm Q})^{2/3}$.

The case of opposite-charge Coulomb scattering in a magnetic field reveals complicated chaotic motion \citep{Hu2002}. 
The transfer of Coulomb energy to gyrational motion allows the classical electron to remain in the neighborhood of the
ion for long intervals, whose precise duration is extremely sensitive to the energy at infinity.   Furthermore, the cross
section tends to zero as $B \rightarrow \infty$, because backscattering requires motion across the magnetic field.  This suppression is absent in the quantum case, because the incident wavepacket can backscatter off an attractive potential.

\begin{figure}
\epsscale{0.8}
\plotone{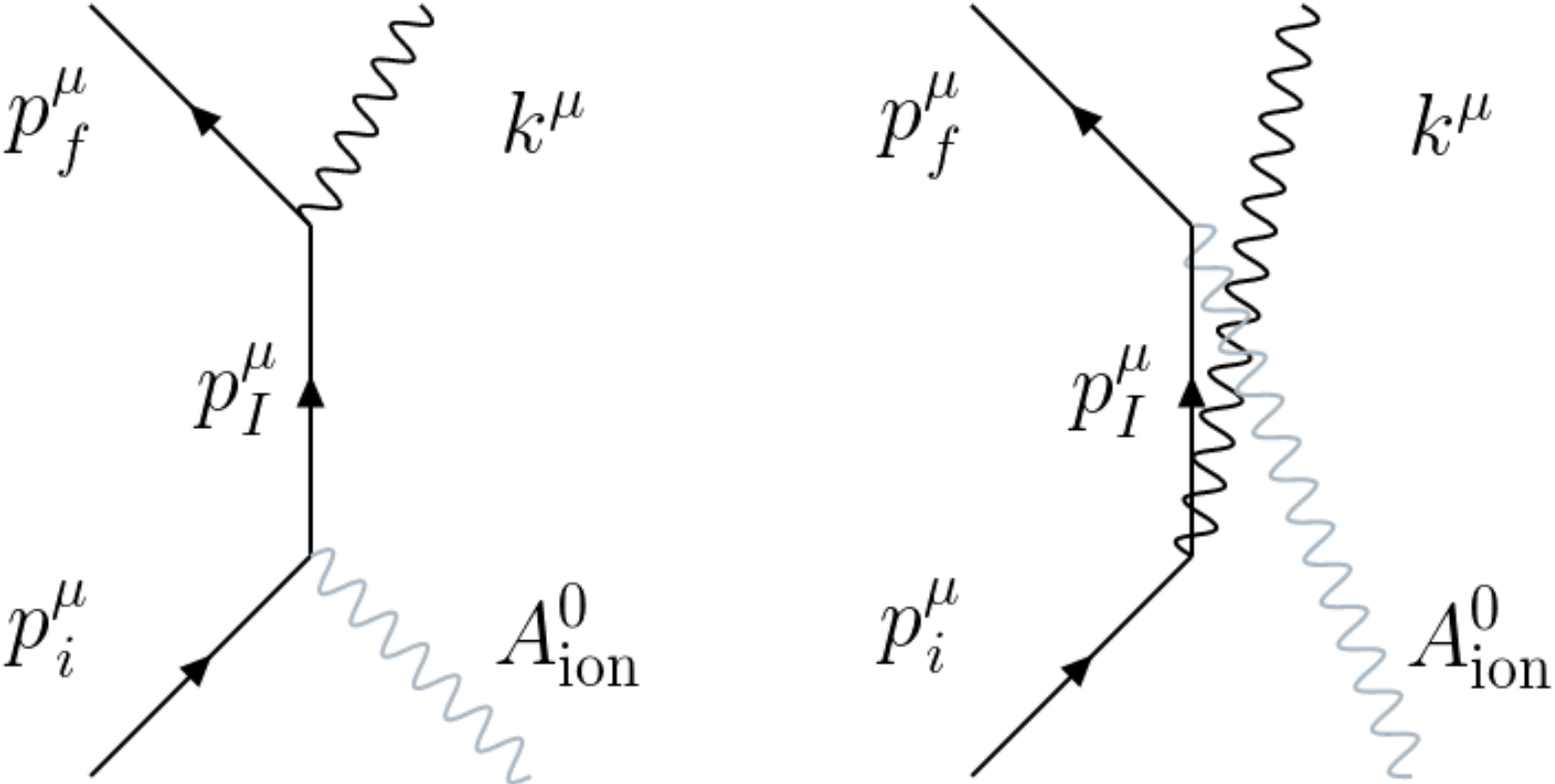}
\caption{Feynman diagrams for bremsstrahlung, evaluated in the approximation where the ion is infinitely heavy and
represented by a fixed Coulomb potential (gray line).}\label{BrShort}
\end{figure}

\section{Relativistic $\MakeLowercase{e}^\pm$-Ion Bremsstrahlung}\label{s:bremss}

We now consider bremsstrahlung (free-free) emission by an electron or positron interacting with a heavy
ion of charge $Ze$, as modified by the presence of a background magnetic field (Figure \ref{BrShort}),
\be
Z+e^-_i \rightarrow Z+e^-_f + \gamma.
\ee
In contrast with \cite{PP1976},
we allow for relativistic electron motion but focus on the case $B \gg B_{\rm Q}$, where the initial, intermediate,
and final electron lines can be restricted to the lowest Landau level. We then calculate the thermally averaged emissivity
for a range of temperatures.  The error introduced by the adoption of the Born approximation (the neglect of the electrostatic
electron-ion interaction in the spinor wave functions) is also quantified:  it decreases with increasing electron temperature. 
We also show that radiative recombination into bound electron-ion states is suppressed relative to free-free emission by a
strong magnetic field.

\subsection {Derivation} \label{BremssDerive}

As in Section \ref{EpmIonSc}, we treat the ion as a fixed Coulomb field centered at $x=y=z=0$ and
neglect gyrational excitation of the ion in the initial state. 
The emission of a photon of energy $\simeq eB/m_p$ has a resonantly enhanced cross section associated
with the transition of the ion to a higher Landau level (e.g. \citealt{PL2007}),  but in a super-QED magnetic field this 
need not exceed the cross section for free-free emission with the ion remaining in the ground Landau state, being
suppressed by a factor $\sim (B/B_{\rm Q})^{-1}$.
For transrelativistic electrons, the relative importance of resonant emission is further reduced by the fact that
most of the energy in free-free photons is radiated near $\omega \sim m$ (when there is no Landau transition of the ion),
well above the $\sim 10$ keV energy of the proton cyclotron line near the surface of a magnetar.\footnote{ For 
subrelativistic electrons, the ratio of resonant to nonresonant bremsstrahlung cross sections is 
$\sim (3\pi/8) (\alpha_{\rm em} B/B_{\rm Q})^{-1}$ at $\omega \simeq eB/m_p$, which is not much different 
from unity when $B \sim 10^2\,B_{\rm Q}$.  This may be checked by considering the inverse process of free-free 
absorption, as given by Equations (B1)-(B11) of \cite{PL2007} and applying Kirchhoff's law.  The resonant contribution
to the cross section (B1) for free-free absorption (polarization index $\alpha = +1$) is averaged over frequency near 
$\omega = eB/m_p$, and compared with the nonresonant contribution (polarization index $\alpha = 0$).
Note that whereas both E-mode and O-mode photons have a component of their electric vectors proportional to
the basis vector $(\varepsilon_x + i\varepsilon_y)/\sqrt{2}$ (corresponding to $\alpha = +1$), only the O-mode overlaps with 
$\varepsilon_z$ ($\alpha = 0$).}

Following the same logic outlined in Section \ref{EpmIonSc},
we assign the incoming electron an impact parameter $a_i$ with respect to the ion.  (The calculation for positron-ion
bremsstrahlung gives an identical result.)  The first term in the matrix element is
\begin{multline}
S_{fi}[1] = - ie^2\int d^4x\int d^4x'\left [ \bar{\psi}_-^{(-1)}(x') \right ]_{p_{z,f},n=0,a_f} 
\gamma_{\nu} A^\nu(x)^* G_f(x'-x)\gamma_{0} A_{\rm ion}^0({\bm x})\left [ \psi_-^{(-1)}(x) \right ]_{p_{z,i},n=0,a_i}.
\end{multline}
Substituting for the electron propagator from Equation (\ref{ComptProp}), this becomes
\begin{equation}\label{eq:sfibremss}
S_{fi}[1] = -\frac{ie^2}{2\pi} \left({L\over 2\omega}\right)^{1/2} \delta(E_i - E_f - \omega)\int dp_I 
\int \frac{da_I}{\lambda_B^2}\left ( \frac{I_1 I_2}{E_{i} -E_I} + \frac{I_3 I_4}{E_{i} +E_I} \right ),
\end{equation}
where 
\begin{equation}
I_1 = \int d^3x \left [ u_{0,a_I}^{(-1)*} ({\bm x})\right ]^T\frac{Ze}{4\pi\sqrt{x^2 + y^2  + z^2}}
u_{0,a_{i}}^{(-1)}({\bm x})e^{i({\bm p}_{i}-{\bm p}_I)\cdot{\bm x}_\perp + i(p_{z,i}-p_{z,I})z}
\end{equation}
and
\be\label{eq:I2}
I_2 = \int d^3x' \left [ u_{0,a_f}^{(-1)*}({\bm x}') \right ]^T\gamma_0\gamma_{\nu} \left (\varepsilon_f^{\nu}
e^{i{\bm k}\cdot{\bm x}'}  \right )^*
u_{0,a_I}^{(-1)}({\bm x}')e^{i(-{\bm p}_f+{\bm p}_I)\cdot{\bm x}'_\perp + i(-p_{z,f}+p_{z,I})z'}.
\ee
The integral $I_3$ is obtained from $I_1$, and $I_4$ from $I_2$, by replacing
$u_{0,a_I}^{(-1)}({\bm x}')$ with the negative-energy wave function $v_{0,a_I}^{(+1)}({\bm x}')$, 
and taking ${\bm p}_I \rightarrow - {\bm p}_I$. 

These integrals are evaluated in Appendix \ref{BremssInteg}; substituting into Equation (\ref{eq:sfibremss}) gives
\ba
S_{fi}[1] &=& \frac{iZe^3 \pi}{(2 \omega L^7)^{1/2}} \,\delta(E_i - E_f -\omega)
\, \delta\left ( \frac{a_{f} - a_{i}}{\lambda_B^2} - k_y \right )  (\varepsilon^z)^*
e^{-ik_x(a_i + a_f)/2} e^{-\lambda_B^2 k_{\perp}^2/4} \nn
&& \quad \quad \times { K_0[(p_{z,i} - p_{z,f} - k_z)a_i] F(p_{z,i},p_{z,f},k_z) 
    \over [E_i E_f (E_i+m)(E_f+m)]^{1/2}(E_{i}^2 - E_I^2)}.
\ea
Here
\begin{equation}
F  \;\equiv\; p_{z,i}(E_f + m)(E_i+ E_f - 2m) + p_{z,f} (E_i + m)(E_i + E_f + 2m) + 
   k_z[p_{z,i} p_{z,f} + (E_i + m)(E_f + m)],
\end{equation}
where $E_I^2 = p_{z,I}^2 + m^2 = (p_{z,f} + k_z)^2 + m^2$, and $E_{i}^2 - E_I^2 = 
(p_{z,i} - p_{z,f} - k_z)(p_{z,i} + p_{z,f} + k_z) \equiv \Delta p_\parallel (p_{z,i} + p_{z,f} + k_z)$.
The absence of a delta function in $p_z$ once again follows from the assumption of an immobile ion.
The second term in the matrix element is related to $S_{fi}[1]$ by substituting $p_{z,i} \leftrightarrow p_{z,f}$
and $k_z\rightarrow -k_z$ (corresponding to $p_{z,I} = p_{z,f} + k_z \rightarrow p_{z,i} - k_z$).  Summing the two terms, we find
\ba
S_{fi}[1] + S_{fi}[2]  &=&  \frac{iZe^3 \pi}{ (2\omega L^7)^{1/2}} \,
 \delta(E_i - E_f - \omega)\, \delta\left ( \frac{a_{f} - a_{i}}{\lambda_B^2} - k_y \right )   
(\varepsilon^z)^*  e^{-ik_x(a_{f} + a_{i})/2} e^{-\lambda_B^2 k_{\perp}^2/4} \nn
&&  \times K_0(\Delta p_\parallel\,a_i)
  { 4m (p_{z,i} + p_{z,f})[p_{z,f}(E_i + m) - p_{z,i} (E_f + m)] \over
   [E_i E_f (E_i+m)(E_f+m)]^{1/2} \Delta p_\parallel[(p_{z,i} + p_{z,f})^2 - k_z^2]}.
\ea

We now take $\lambda_B \rightarrow 0$ and calculate the differential cross section for a beam of particles:
\be
{1\over 2\pi a_i}{d\sigma \over da_i} = \frac{1}{|\beta_{i}|}\frac{L}{T}\int L^3\frac{\omega^2 d \omega d\Omega}{(2\pi)^3}
\int \frac{L da_{f}}{2 \pi \lambda_B^2}\int L \frac{dp_{z,f}}{2 \pi}\bigl| S_{fi} [1] + S_{fi}[2] \bigr|^2.
\ee
Integrating over the cross section of the beam and using the identity $\int_0^\infty t K_0^2(t) dt = {1\over 2}$ gives
\be\label{MagBremssSig}
\omega \frac{d^2 \sigma}{d \omega d\Omega} =  \sum_{p_{z,f}}
\frac{4Z^2}{\pi}\left ( \frac{e^2}{4 \pi} \right )^3\frac{\left |\varepsilon_z  \right |^2}{|\beta_{i} \beta_{f}|} 
\frac{\omega^2 m^2(p_{z,i} + p_{z,f})^2[p_{z,f}(E_{i} +m)-p_{z,i}(E_{f}+m)]^2}
{(\Delta p_{\vert \vert})^4[(p_{z,i}+p_{z,f})^2-k_z^2]^2E_{i} E_{f}(E_{i}+m)(E_{f}+m)}.
\ee
Here we sum over the two energetically permissible values of the final momentum, $p_{z,f} = \pm \sqrt{E_f^2-m^2}$.

Note that the integral over impact parameter is convergent, due to the presence of a strong background magnetic field:
the kinetic momentum of the outgoing electron is directed along the magnetic field.
The integral of Equation (\ref{MagBremssSig}) over the solid angle is elementary but cumbersome;  it is used in the
numerical evaluation of free-free emission from a thermal plasma but is not repeated here.

The effect of Debye screening of the ion's Coulomb field is easy to include in this calculation, by multiplying
the electrostatic potential by $\exp(-r/r_D)$.  However, in many instances screening has a negligible effect.
The dominant contribution in the integral over impact parameter comes from 
$a_i \sim  1/\Delta p_\parallel \propto \omega^{-1}$.   On the other hand, the Debye screening length is proportional to the
inverse of the plasma frequency.  This means that screening can be ignored as long as $\omega$ lies well 
above the plasma cutoff, because $r_D$ is much larger than the dominant emission impact parameter.

\begin{figure}
\epsscale{0.85}
\plotone{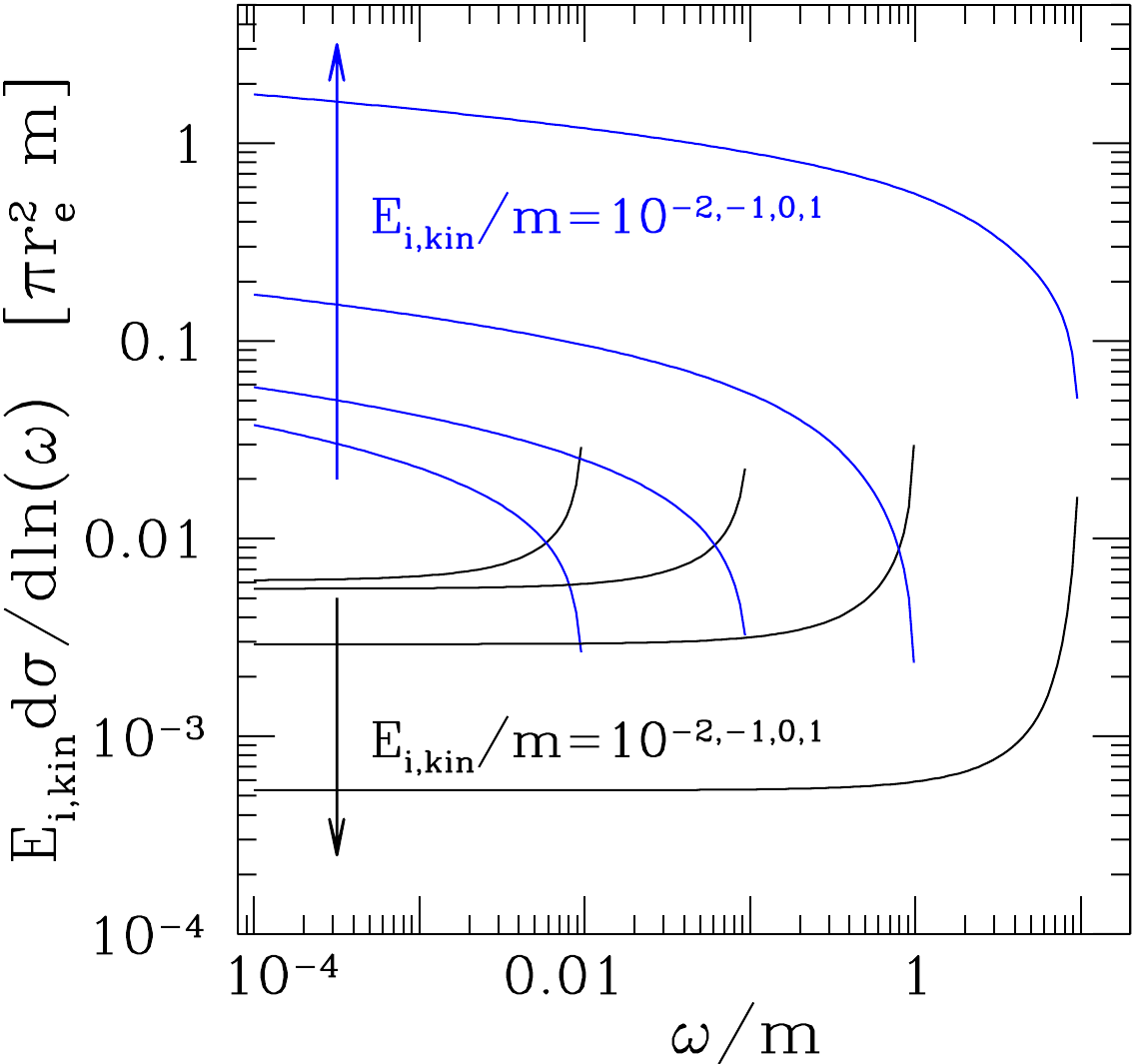}
\caption{Comparison of the relativistic bremsstrahlung cross section for electrons interacting with
protons ($Z = 1$) in a background magnetic field (integral of Equation (\ref{MagBremssSig}) over solid angle; 
black curves) with the relativistic bremsstrahlung cross section in free space (\citealt{BH1934}; blue curves).
Sequences of curves correspond to a range of initial kinetic energy $E_{i,{\rm kin}} = (\gamma_i-1)m$. The cross sections
have been multiplied by a factor of $E_{i, \rm kin}$ for clarity.   Each curve is cut off at the maximum photon 
frequency $\omega = E_{i,\rm kin}$.}\label{BetheCompare}
\end{figure}

We next compare the integral of Equation (\ref{MagBremssSig}) over solid angle with the formula derived by 
\cite{BH1934} for relativistic free-free emission in the Born approximation and at $B = 0$:
\ba\label{BetheSigma}
\omega\frac{d \sigma}{d \omega} &=& {Z^2\over m^2}\left ( \frac{e^2}{4 \pi} \right )^3 \frac{p_f}{p_i}
\Biggl\{ \frac{4}{3} - 2 E_i E_f \frac{p_i^2 + p_f^2}{p_i^2 p_f^2} + m^2 
\left ( \frac{\epsilon_i E_f}{p_i^3} + \frac{\epsilon_f E_i}{p_f^3} - \frac{\epsilon_i \epsilon_f}{p_i p_f} \right ) \nn
&+& L_g\left [ \frac{8 E_i E_f}{3 p_i p_f} + \frac{\omega^2 (E_i^2 E_f^2 + p_i^2 p_f^2)}{p_i^3 p_f^3} \right ]  + 
\frac{m^2 \omega L_g}{2 p_i p_f} 
\left [ \frac{E_i E_f + p_i^2}{p_i^3} \epsilon_i - \frac{E_i E_f + p_f^2}{p_f^3}\epsilon_f + 
\frac{2 \omega E_i E_f}{p_i^2 p_f^2} \right ] \Biggr\}.
\ea
Here
\begin {equation}
\epsilon_{i,f} = 2 \ln \frac{E_{i,f} + p_{i,f}}{m}; \quad\quad L_g = 2 \ln\frac{E_i E_f + p_i p_f - m^2}{m \omega}.
\end {equation}
Figure \ref{BetheCompare} shows that the magnetic cross section is generally smaller than Equation (\ref{BetheSigma}),
except near the limiting frequency $E_i-m$ where the final-state electron moves slowly.

It is possible to improve on the Born approximation for relativistic bremsstrahlung by correcting 
the electron wave function for the electrostatic interaction with the ion \citep{EH1969, NIK1998, VH2015}.
In the absence of a background magnetic field, this produces the correction factor first derived by
\cite{Elwert1939}.  Obtaining a similar correction factor for the magnetic case is 
beyond the scope of the present investigation.  Nonetheless, Equation (\ref{MagBremssSig}) is
expected to be a good approximation as long as the kinetic energy of the final-state electron (or positron) exceeds 
the hydrogen binding energy (Equation (\ref{MagHBind})); see Section \ref{s:born}.

\subsection {Limiting Cases}

The bremsstrahlung cross section (Equation (\ref{MagBremssSig})) simplifies in various regimes.
At low emission frequencies, $\omega \ll E_{i,\rm kin} = E_i - m$, one has
\be
\omega\frac{d^2 \sigma}{d \omega d\Omega}  = |\varepsilon_z|^2\, Z^2\alpha_{\rm em} r_e^2 
{ (m/E_i)^4 \over \pi \beta_{i}^2 (1-\beta_{i} \cos\theta)^2 } 
\left[ {(m/E_i)^2\over (1-\beta_i\cos\theta)^2} + {1\over (1+\beta_i\cos\theta)^2}\right ].
\ee
The two terms on the right-hand side represent emission by forward-scattered and back-scattered $e^\pm$.
The second term can be obtained from the Coulomb backscattering cross section (Equation (\ref{ElIonSigma0})) by
multiplying by the soft photon factor $ \alpha_{\rm em} (\omega/2\pi)^2 (\varepsilon\cdot p_f / k\cdot p_f -
\varepsilon\cdot p_i / k\cdot p_i)^2$. When the incident electron moves subrelativistically,
\begin {equation}
\omega\frac{d^2 \sigma}{d \omega d\Omega}  = |\varepsilon_z|^2\,Z^2\alpha_{\rm em} r_e^2
\frac{2}{\pi |\beta_i \beta_f|}.
\end {equation}

\begin{figure}
\epsscale{0.85}
\plotone{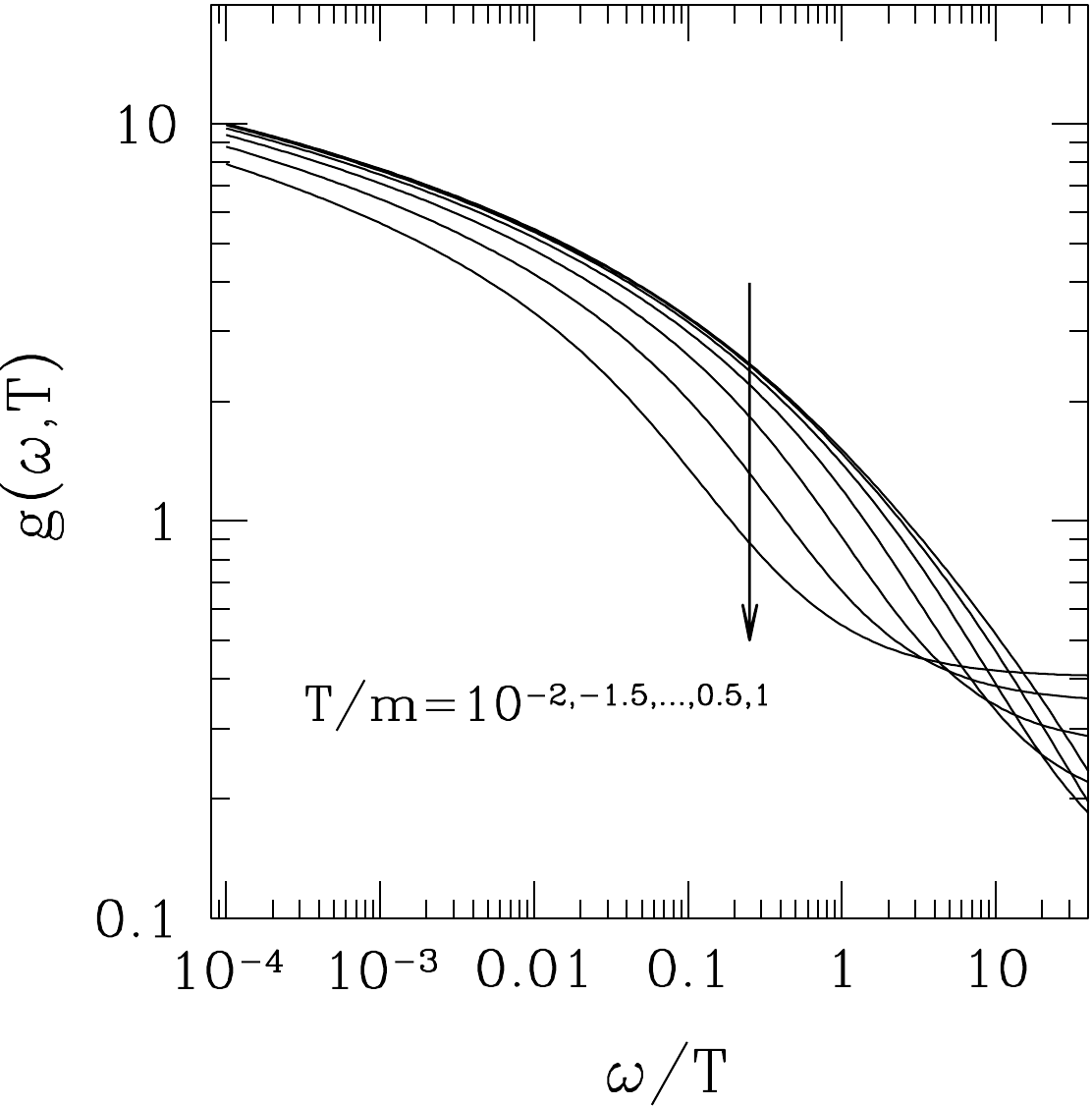}
\caption{Gaunt factor as a function of $\omega/T$ for a range of temperatures $T$.}
\label{GauntFig}
\end{figure}

\subsection{Thermal Bremsstrahlung} \label{ThermBremss}

We now evaluate the free-free emission from a thermal electron-proton plasma ($Z = 1$), 
with electrons confined to the lowest Landau level,
\begin {equation}
\frac{d^2n_\gamma}{d \omega dt} = n_e n_p \left\langle |\beta_i|\frac{d\sigma}{d\omega} \right\rangle.
\end {equation}
Here the cross section (Equation (\ref{MagBremssSig})) has been integrated over the solid angle of the emitted
photon; $n_e$, $n_p$, and $n_\gamma$ are the number densities of electrons, protons, and photons, respectively;  and
\begin {equation} \label{ThermAvDefine}
\left \langle |\beta_i|\frac{d\sigma}{d\omega} \right \rangle = 
\frac{ \int_{p_{z, \rm min}}^{\infty} dp_{z,i} \beta_{i} \exp[-\gamma_i m/T]\,d\sigma/d\omega}
{\int_{0}^{\infty} dp_{z,i} \exp[-\gamma_i m/T]}.
\end {equation}
The cutoff $p_{z, \rm min} = \sqrt{\omega^2+2m\omega}$ corresponds to an incident electron of the
minimum energy needed to emit a photon of energy $\omega$.  We follow convention and describe this 
thermal average in terms of a Gaunt factor $g(T,\omega)$, defined as
\begin {equation} \label{BremssGauntDef}
\left\langle |\beta_i| \frac{d\sigma}{d\omega} \right\rangle  = 
\frac{8\alpha_{\rm em}r_e^2}{3\omega} \frac{e^{-(m + \omega)/T}}{K_1(m/T)}g(T,\omega).
\end {equation}
This integral over the momentum of the incoming electron is performed numerically.
The result is shown in Figure \ref{GauntFig} for a range of temperatures and is tabulated in 
Table \ref{GauntTable}.

The Gaunt factor can be evaluated analytically in some limiting cases.  When $m \gg T \gg \omega$,
\begin {equation}\label{LowOmGaunt}
g(T,\omega) = \ln \left (\frac{4T}{\omega} \right) - \gamma_{\rm EM},
\end {equation}
where $\gamma_{\rm EM}$ is the Euler-Mascheroni constant.  
Next, when $m \gg \omega \gg T$, we have
\begin {equation}\label{LowTMidOmGaunt}
g(T,\omega) = \sqrt{\frac{\pi T}{\omega}}.
\end {equation}
Both of these expressions agree with the nonrelativistic results of \cite{PP1976} in the
regime $B \gg B_{\rm Q}$ (compare their Equations (38) and (41)).  
Equations (\ref{LowOmGaunt}) and (\ref{LowTMidOmGaunt}) resemble the results for free-free
emission from an unmagnetized plasma with $m \gg T \gg 1$ Ry, for which
$g(T,\omega) \simeq (\sqrt{3}/\pi) [\ln(4T/\omega) - \gamma_{\rm EM}]$ when $\omega \ll T$, and 
$g(T,\omega) \simeq (3T/\pi\omega)^{1/2}$ when $\omega \gg T$ (e.g. \citealt{NT1973}).   

\begin{longtable}[l]{l c c c c c c c c c c c c c c c c}
\caption{Gaunt Factor and Related Error for $T/m = 10^{-2,-1.5,...,0.5,1}$ and $B = 100\,B_{\rm Q}$} \label{GauntTable}\\
\hline
& \multicolumn{2}{c}{\footnotesize $T/m = 10^{-2}$} & \multicolumn{2}{c}{\footnotesize$T/m = 10^{-1.5}$}&\multicolumn{2}{c}{\footnotesize$T/m = 10^{-1}$} & \multicolumn{2}{c}{\footnotesize$T/m = 10^{-0.5}$} & \multicolumn{2}{c}{\footnotesize$T/m = 10^0$} & \multicolumn{2}{c}{\footnotesize$T/m = 10^{0.5}$}& \multicolumn{2}{c}{\footnotesize$T/m = 10$} \\
 \cline{2-17}
$\omega / m$ & $g$ & $\delta g$ & $g$ & $\delta g$ & $g$ & $\delta g$ & $g$ & $\delta g$ & $g$ & $\delta g$ & $g$ & $\delta g$ & $g$ & $\delta g$ \\
 \hline
\endfirsthead

 \multicolumn{17}{c}{Table \ref{GauntTable} continued} \\
 \hline
& \multicolumn{2}{c}{ \footnotesize$T/m = 10^{-2}$} & \multicolumn{2}{c}{\footnotesize$T/m = 10^{-1.5}$}&\multicolumn{2}{c}{\footnotesize$T/m = 10^{-1}$} & \multicolumn{2}{c}{\footnotesize$T/m = 10^{-0.5}$} & \multicolumn{2}{c}{\footnotesize$T/m = 10^0$} & \multicolumn{2}{c}{\footnotesize$T/m = 10^{0.5}$}& \multicolumn{2}{c}{\footnotesize$T/m = 10$}  \\
 \cline{2-17}
$\omega / m$ & $g$ & $\delta g$ & $g$ & $\delta g$ & $g$ & $\delta g$ & $g$ & $\delta g$ & $g$ & $\delta g$ & $g$ & $\delta g$ & $g$ & $\delta g$ \\
 \hline
 \endhead 

 \hline
 \endfoot

 \endlastfoot

\fontsize{9.5}{11}\selectfont0.0100 &  \fontsize{9.5}{11}\selectfont1.5093 &  \fontsize{9.5}{11}\selectfont0.7770 &  \fontsize{9.5}{11}\selectfont2.2798 &  \fontsize{9.5}{11}\selectfont0.8081 &  \fontsize{9.5}{11}\selectfont3.1692 &  \fontsize{9.5}{11}\selectfont0.8184 &  \fontsize{9.5}{11}\selectfont4.0545 &  \fontsize{9.5}{11}\selectfont0.8216 &  \fontsize{9.5}{11}\selectfont4.7988 &  \fontsize{9.5}{11}\selectfont0.8227 &  \fontsize{9.5}{11}\selectfont5.3180 &  \fontsize{9.5}{11}\selectfont0.8230 &  \fontsize{9.5}{11}\selectfont5.6217 &  \fontsize{9.5}{11}\selectfont0.8231\\
 \fontsize{9.5}{11}\selectfont0.0107 &  \fontsize{9.5}{11}\selectfont1.4678 &  \fontsize{9.5}{11}\selectfont0.7517 &  \fontsize{9.5}{11}\selectfont2.2267 &  \fontsize{9.5}{11}\selectfont0.7817 &  \fontsize{9.5}{11}\selectfont3.1081 &  \fontsize{9.5}{11}\selectfont0.7917 &  \fontsize{9.5}{11}\selectfont3.9889 &  \fontsize{9.5}{11}\selectfont0.7949 &  \fontsize{9.5}{11}\selectfont4.7311 &  \fontsize{9.5}{11}\selectfont0.7959 &  \fontsize{9.5}{11}\selectfont5.2494 &  \fontsize{9.5}{11}\selectfont0.7962 &  \fontsize{9.5}{11}\selectfont5.5529 &  \fontsize{9.5}{11}\selectfont0.7963\\
 \fontsize{9.5}{11}\selectfont0.0115 &  \fontsize{9.5}{11}\selectfont1.4270 &  \fontsize{9.5}{11}\selectfont0.7270 &  \fontsize{9.5}{11}\selectfont2.1742 &  \fontsize{9.5}{11}\selectfont0.7561 &  \fontsize{9.5}{11}\selectfont3.0473 &  \fontsize{9.5}{11}\selectfont0.7658 &  \fontsize{9.5}{11}\selectfont3.9235 &  \fontsize{9.5}{11}\selectfont0.7688 &  \fontsize{9.5}{11}\selectfont4.6635 &  \fontsize{9.5}{11}\selectfont0.7698 &  \fontsize{9.5}{11}\selectfont5.1809 &  \fontsize{9.5}{11}\selectfont0.7701 &  \fontsize{9.5}{11}\selectfont5.4840 &  \fontsize{9.5}{11}\selectfont0.7703\\
\hline
\end{longtable}

\subsection{Validity of the Born Approximation}\label{s:born}

Equation (\ref{MagBremssSig}) for the cross section, which is obtained in the Born approximation,
is a good first-order approximation only as long as the kinetic energy of
the final-state electron (or positron) exceeds the hydrogen binding energy in a magnetic field (\ref{MagHBind}), $E_f-m  \gg 
|E_{\rm H,0}|$.  However, as can be seen from Equation (\ref{MagBremssSig}) and from Figure \ref{BetheCompare}, the cross 
section diverges as $\beta_f \rightarrow 0$.  Although the thermal average of $d\sigma/d\omega$ 
is convergent, a significant portion of the integral in Equation (\ref{ThermAvDefine}) may
be supplied by low momenta where the electrostatic correction to the electron wave function is significant.

To quantify this uncertainty, we have separately evaluated the part $\delta g$ of the momentum integral in
Equation (\ref{ThermAvDefine}) supplied by $p_{z,\rm min} \leq p_{z,i} \leq \sqrt{|E_{\rm H,0}|(2m+|E_{\rm H,0}|)}$.   
This estimate of the error in $g(\omega,T)$ is included in Table \ref{GauntTable}.  As the temperature becomes more
relativistic, the fractional error $\delta g/g$ drops.   The uncertainty is relatively large at subrelativistic
temperatures.  We expect this result to persist in weaker magnetic fields even when the emissivity is calculated
to all orders in the magnetic field, as was done by \cite{PP1976}.

\subsection{Comparison with Radiative Recombination}\label{s:recomb}

In an unmagnetized plasma, the radiative recombination of hydrogen has a relatively large cross section
compared with free-free emission when the kinetic energy of the unbound electron is comparable to the ionization
energy.  The radiative free-bound cross section scales as $\sigma_{\rm fb} \sim r_e^2/\alpha_{\rm em}$ as compared
with $\sim \alpha_{\rm em} r_e^2$ for free-free emission.   Although these scalings with $\alpha_{\rm em}$ persist
in the presence of an ultrastrong magnetic field, as we now show, other factors emerge that suppress free-bound
emission relative to free-free emission.  The net result is that free-bound emission is only competitive for $\omega > T$.   

The cross section $\sigma_{\rm fb}$ may be obtained from the photoionization cross section $\sigma_{\rm bf}$
using the principle of detailed balance.  We consider only subrelativistic electrons and photons
of energy $\omega \ll m$ interacting with free protons and neutral hydrogen, in which case \citep{GPT1974}
\begin {equation} \label{IonizationSigma}
\sigma_{\rm bf} = {8 \pi \alpha_{\rm em}\over m\omega}
\left ( \frac{|E_{\rm H,0}|}{\omega}  \right )^{3/2}
\left( 2 \sin^2\theta+ \frac{\omega}{2 \omega_B} \cos^2 \theta \right).
\end {equation}
Here $\theta$ is the propagation angle of the absorbed photon with respect to ${\bm B}$ and $\omega_B \equiv eB/m$.
This cross section represents recombination to the most tightly bound hydrogen energy level, with angular 
momentum quantum number $s=0$.

To obtain the free-bound cross section, we focus on the simplest case of a partially ionized hydrogen gas interacting
with a blackbody radiation field of a low enough temperature that cyclotron excitations of the protons can be neglected.
The densities of electrons and photons are
\be
{dn_e\over dp_z} = n_e {e^{-p_z^2/2mT}\over (2\pi m T)^{1/2}}; \quad\quad
 {d^2n_\gamma\over d\omega d(\cos\theta)} = {\omega^2\over (2\pi)^2}N_\gamma,
\ee
where $N_\gamma = 1/(e^{\omega/T}-1)$.
Detailed balance implies the following relation between the photoionization and recombination cross sections:
\be \label{RecombRateBalance}
(1+N_\gamma){p_z\over m} \frac{d \sigma_{\rm fb}}{d(\cos\theta)} n_p {dn_e\over dp_z} dp_z
= \sigma_{\rm bf} n_{\rm H} \frac{d^2n_\gamma}{d \omega d(\cos\theta)} d\omega.
\ee
The ratio $n_p n_e / n_{\rm H}$, where $n_{\rm H}$ is the density of neutral H atoms, can be obtained by considering
the Boltzmann law, as applied to a two-level system consisting of a neutral atom and a proton paired with a free
electron moving in the momentum range $(p_z,p_z+dp_z)$:
\be \label{RecombBoltzmann}
\frac{dn_p(p_z)}{n_{\rm H}} = \frac{g_pg_e}{g_H} \exp\left[ -{ |E_{\rm H,0}| + p_z^2/2m \over T} \right].
\ee
Here the $g_i$ label the quantum degeneracies of the various states ($g_{\rm H} = g_p = 1$) and
\be
g_e = \frac{eB dp_z}{n_e (2 \pi)^2}
\ee
is the differential statistical weight of the electron.  Substituting these relations into Equation 
(\ref{RecombBoltzmann}) and integrating over $p_z$, we obtain the relevant Saha ionization relation,
\be
\frac{n_p n_e}{n_{\rm H}} = \frac{e B}{(2 \pi)^{3/2}} (mT)^{1/2} \exp\left [ -\frac{|E_{\rm H,0}|}{T} \right ].
\ee
Substituting this into Equation (\ref{RecombRateBalance}) yields the generalized Milne relation
\be\label{eq:milne}
\frac{d \sigma_{\rm fb}}{d(\cos\theta)} = \sigma_{\rm bf} \frac{\omega^2}{e B}.
\ee
As required, this relation is independent of collective properties such as temperature.
Finally, substituting Equation (\ref{IonizationSigma}) gives the differential cross section,
\be
\frac{d \sigma_{\rm fb}}{d(\cos\theta)} 
= 8 \pi {r_e^2\over\alpha_{\rm em}}  \left(\frac{\omega m}{e B}\right)  \left ( \frac{|E_{\rm H,0}|}{\omega}  \right )^{3/2}
\left( 2 \sin^2\theta +\frac{\omega}{2 \omega_B} \cos^2 \theta \right).
\ee
Integrating over the photon direction gives
\be
\sigma_{\rm fb}= 16 \pi {r_e^2\over\alpha_{\rm em}}\,\left({\omega/m\over B/B_{\rm Q}}\right)
 \left ( \frac{|E_{\rm H,0}|}{\omega}  \right )^{3/2}\left( \frac{4}{3} +\frac{\omega}{6 \omega_B} \right).
\ee
In the nonrelativistic regime considered here, this cross section is significantly smaller than that of electron-ion backscattering, as given by Equation (\ref{ElIonSigma}).

\begin{figure}
\epsscale{0.85}
\plotone{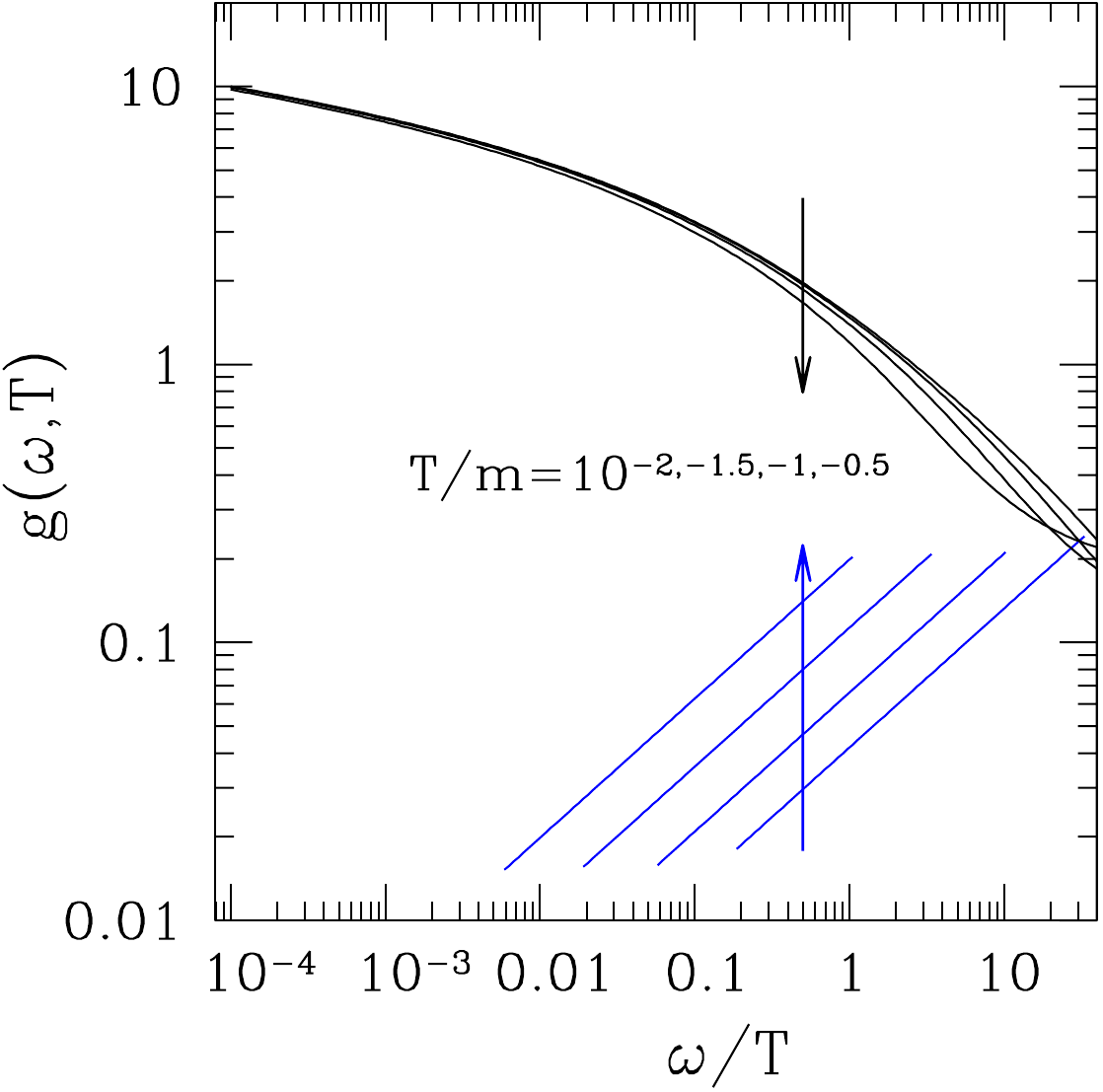}
\caption{Comparison of the bremsstrahlung Gaunt factor $g(T,\omega)$ (black curves) and the recombination Gaunt factor $g_R(T,\omega)$ (blue curves) for a range of subrelativistic temperatures and $B = 100 B_{\rm Q}$.  The arrows
point in the direction of increasing $T$.  Free-bound emission is significantly weaker than free-free emission 
except at high frequencies, where the emission is exponentially suppressed.}\label{RecombGauntCompare}
\end{figure}

In order to compare the relative importance of recombination emission and bremsstrahlung, we consider the emission
rate from a thermal distribution of electrons,
\be
n_e n_p  \left\langle |\beta| \frac{d \sigma_{\rm fb}}{d \omega}  \right\rangle
\equiv 2{p_\omega\over m} \sigma_{\rm fb} n_p \frac{dn_e}{dp_z}\biggl|_{p_z=p_\omega}
\frac{dp_z}{d\omega}\biggr|_{p_z=p_\omega},
\ee
where $p_\omega \equiv \sqrt{2m(\omega - |E_{\rm H,0}|)}$, and $dp_z/d\omega = m/p_\omega$ when evaluated at
$p_z=p_\omega$.  Defining a recombination Gaunt factor $g_R(T,\omega)$ similarly to Equation (\ref{BremssGauntDef}),
one has
\begin {equation} \label{RecombGaunt}
g_R(T,\omega) = \frac{6\pi \omega^2}{\alpha_{\rm em}^2 e B}\left ( \frac{|E_{\rm H,0}|}{\omega} \right )^{3/2} 
\exp\left[{|E_{\rm H,0}|\over T}\right]
\left(\frac{4}{3} + \frac{\omega}{6 \omega_B} \right)\quad\quad(\omega \geq |E_{\rm H,0}|).
\end {equation}
Figure \ref{RecombGauntCompare} plots both this expression and $g(T,\omega)$ over a range of frequencies and 
temperatures.

\section{Summary}

Continuing the approach of \cite{KT2018}, we have derived and analyzed the rates of several electromagnetic processes in the
presence of an intense magnetic field: electron-positron scattering, $e^\pm$-ion scattering, and $e^\pm$-ion bremsstrahlung.
The restriction of real and virtual $e^\pm$ to the lowest Landau state allowed us to obtain relatively compact
closed-form expressions for cross sections.

A summary of our main results follows:

1. The cross section for the scattering of very strongly magnetized electrons and positrons is calculated for the first time
(Section \ref{BhabhaSection}).  
This cross section is regulated by the finite width of intermediate-state photons with energy exceeding $2m$
(the second term in Equation (\ref{BhabhaSigmaFinal})).  

2. The calculation of $e^\pm$-ion backscattering is generalized to include both relativistic motion and Debye screening 
(Section \ref{EpmIonSc} and Equation (\ref{ElIonSigma})), with the ion treated as immobile.

3. The cross section for relativistic bremsstrahlung is derived in Section \ref{s:bremss} in the Born approximation, 
with the ion treated as immobile, and compared with the nonmagnetic result \citep{BH1934}.  At low frequencies 
and particle energies, the magnetic field suppresses the cross section by about an order of magnitude, with the 
suppression becoming stronger at higher energies (Equation (\ref{MagBremssSig}) and Figure \ref{BetheCompare}).

4. The free-free emission is averaged over a one-dimensional thermal electron distribution.  The corresponding 
Gaunt factor is tabulated in Table \ref{GauntTable}, and analytic nonrelativistic approximations to it are derived.
The uncertainty in the thermal bremsstrahlung emissivity arising from the neglect of the electron-ion interaction
in the spinor wave functions is quantified (Section \ref{s:born}).  This uncertainty becomes proportionately smaller as the
electrons grow more relativistic.

5. The rate of photon emission by radiative recombination is derived and compared with thermal bremsstrahlung
(Section \ref{s:recomb}).  Free-bound emission is shown to be negligible compared with free-free, except at high
frequencies, $\omega \gg T$. 

\acknowledgements
This work was supported by the NSERC of Canada.  Alexander Kostenko thanks NSERC for graduate fellowship support.
We thank the referee for comments on the presentation of our results.

\appendix

\section {Electron-Positron Scattering Integrals} \label{BhabhaAppendix}

Here we evaluate the matrix element for electron-positron scattering in a general longitudinal reference frame,
as given by Equation (\ref{BhabhaMEInt}).  This involves the combination of integrals $I_{2,\mu} \eta^{\mu \nu}I_{1,\nu}$.
As a first step, note that
\ba
\left [ v_{0,a_{+,i}}^{(+1)*} \right ]^T\gamma_0\gamma_{\nu} \eta^{\mu \nu}u_{0,a_{-,i}}^{(-1)} &=& 
i\frac{\phi_0(x-a_{+,i}) \phi_0(x-a_{-,i})}{2[E_{+,i}E_{-,i}(E_{+,i}+m)(E_{-,i}+m)]^{1/2}} \nn 
&\times&  \left\{p_{z+,i}[\eta^{\mu t}(E_{-,i}+m)-\eta^{\mu z}p_{z-,i}]+
     (E_{+,i}+m)[-\eta^{\mu z}(E_{-,i}+m)+\eta^{\mu t}p_{z-,i}]\right\}.\nn
\ea
Only the $\mu=t$ and $z$ components of $I_{2,\mu}$ contribute; these are, respectively,
\be
\left [ u_{0,a_{-,f}}^{(-1)*}\right ]^T\,v_{0,a_{+,f}}^{(+1)} = -i \frac{p_{z+,f}(E_{-,f}+m)+p_{z-,f}(E_{+,f}+m)}
{2[E_{+,f}E_{-,f}(E_{+,f}+m)(E_{-,f}+m)]^{1/2}} \, \phi_0(x-a_{+,f}) \phi_0(x-a_{-,f});
\ee
and
\be
\left [ u_{0,a_{-,f}}^{(-1)*}\right ]^T\gamma_0\gamma_{z}v_{0,a_{+,f}}^{(+1)} = i\frac{p_{z+,f}p_{z-,f}+(E_{+,f}+m)(E_{-,f}+m)}
{2[E_{+,f}E_{-,f}(E_{+,f}+m)(E_{-,f}+m)]^{1/2}} \, \phi_0(x-a_{+,f}) \phi_0(x-a_{-,f}).
\ee
Substituting these results into Equations (\ref{I1Bhabha}) and (\ref{I2Bhabha}), and making use of the overlap integral in Equation 
(\ref{eq:phiint}), we find
\ba\label{eq:isum}
I_{2,\mu} \eta^{\mu \nu}I_{1,\nu} &=& \frac{ e^{iq_x (a_{+,f} + a_{-,f} - a_{+,i} - a_{-,i})/2} 
e^{-\lambda_B^2(q_x^2 + q_y^2)/2}\, F(p_{z-,i},p_{z+,i},p_{z-,f},p_{z+,f}) } 
{ 4L^4\sqrt{E_{+,i} E_{-,i} E_{+,f} E_{-,f} (E_{+,i}+m)(E_{-,i}+m)(E_{+,f}+m)(E_{-,f}+m)}}  \nn
&\times&  (2\pi)^4 \delta \left( \frac{a_{+,i}-a_{-,i}}{\lambda_B^2} - q_y \right)
\, \delta \left(q_y - \frac{a_{+,f}-a_{-,f}}{\lambda_B^2} \right) \,
 \delta(p_{z+,i} + p_{z-,i} - q_z) \, \delta(q_z - p_{z+,f} - p_{z-,f}),\nn
\ea
where
\ba
F &\;\equiv\;&  -\left[p_{z+,i}p_{z-,i} + (E_{+,i}+m)(E_{-,i}+m)][p_{z+,f}p_{z-,f} + (E_{+,f}+m)(E_{-,f}+m)\right] \nn
&+& \left[p_{z+,i}(E_{-,i}+m) + p_{z-,i}(E_{+,i}+m)][p_{z+,f}(E_{-,f}+m) + p_{z-,f}(E_{+,f}+m)\right].\nn
\ea
Substituting Equation (\ref{eq:isum}) into Equation (\ref{BhabhaMEInt}) gives
\ba
S_{fi}[2] &=& \frac{-i e^2}{4L^4} \int dq_x
\frac{ e^{iq_x (a_{+,f} + a_{-,f} - a_{+,i} - a_{-,i})/2} e^{-\lambda_B^2(q_x^2 + q_y^2)/2} }
{ (E_{+,i} + E_{-,i})^2 - {\bm q}^2 } \nn
&\times& {(2\pi)^2 \delta^{(3)}_{fi}(E,p_y,p_z) F(p_{z-,i},p_{z+,i},p_{z-,f},p_{z+,f}) 
\over \sqrt{E_{+,i} E_{-,i} E_{+,f} E_{-,f} (E_{+,i}+m)(E_{-,i}+m)(E_{+,f}+m)(E_{-,f}+m)}},
\ea
where ${\bm q}^2 = q_x^2 + q_y^2 + q_z^2$, $q_z = p_{z+,i} + p_{z-,i}$ and $q_y = (a_{+,i} - a_{-,i})/\lambda_B^2$. 

\section {Ionic Integrals}\label {BremssInteg}

The matrix elements for both $e^\pm$-ion scattering and $e^\pm$-ion bremsstrahlung contain integrals over the
ion Coulomb field.   The $y$ integral in Equation (\ref{CoulombSpatialIntegral}) works out to
\be
\int_{-\infty}^{\infty} dy \exp \left [ iy \frac{a_i - a_f}{\lambda_B^2} \right ]
\frac{1}{\sqrt{x^2 + y^2 + z^2}}  = 2 K_0 \left [  \frac{(a_i - a_f)\sqrt{x^2 + z^2}}{\lambda_B^2} \right ],
\ee
where $K_0$ is the modified Bessel function.   Taking $\lambda_B \rightarrow 0$, we see that 
this quantity peaks very strongly at $a_i \approx  a_f$.  It may be expressed in terms of a delta function
in $a_f$. Performing the integral
\be
\int_{-\infty}^{\infty} da_f K_0\left [\frac{(a_i - a_f)\sqrt{x^2 + z^2}}{\lambda_B^2} \right ] = 
\frac{\pi \lambda_B^2}{\sqrt{x^2 + z^2}},
\ee
we have
\be
2K_0 \left [  \frac{(a_i - a_f)\sqrt{x^2 + z^2}}{\lambda_B^2} \right ]  \rightarrow 
2\pi\delta\left ( \frac{a_{i} - a_f}{\lambda_B^2} \right ) \frac{1}{\sqrt { x^2 +  z^2}}.
\ee
Given a typical impact parameter $a_i \sim 1/p_{z,i}$, we deduce that $|a_f - a_i| \sim \lambda_B^2 p_{z,i}$,
which is much smaller than the width of the spinor wave functions ($\delta x \sim \lambda_B$).
Hence, we may take $a_i = a_f$ and $x = a_i$ in evaluating the $x$ and $z$ integrals, the latter of which becomes
\be
\int_{-\infty}^{\infty} dz ~ \cos[(p_{z,i} - p_{z,f})z]\frac{1}{\sqrt{a_i^2 + z^2}}  
= 2 K_0[(p_{z,i} - p_{z,f})a_i].
\ee
The Coulomb field is essentially constant over the domain of support of the $x$ integral, which
is easily expressed in terms of the orthogonality relation
\be
\int dx\,\phi_{n_f}(x-a_i)\phi_0(x-a_i) = {\delta_{n_f,0}\over L^2}.
\ee
Along with considerations of energy conservation, this requires the final-state particle to be confined
to the lowest Landau level.   The net overlap of the ingoing and outgoing spinors with the Coulomb potential 
is presented in Equation (\ref{CoulombSpatialIntegral}).

The integrals $I_i$ appearing in the bremsstrahlung matrix element (\ref{eq:sfibremss}) are evaluated in a similar manner,
\be
I_1  = {e\over 4\pi} K_0[(p_{z,i} - p_{z,I})a_i] 
\frac{p_{z,i} p_{z,I} + (E_i + m)(E_I + m)}{L^2[E_i E_I (E_i + m)(E_I + m)]^{1/2}}
\, 2\pi\delta\left ( \frac{a_I - a_i}{\lambda_B^2} \right ),
\ee
and
\be
I_3 = i {e\over 4\pi} K_0[(p_{z,i} + p_{z,I})a_i]
\frac{p_{z,i}(E_I + m) + p_{z,I} (E_i + m)}{L^2[E_i E_I (E_i + m)(E_I + m)]^{1/2}}
\, 2\pi \delta\left ( \frac{a_I - a_i}{\lambda_B^2} \right).
\ee
Meanwhile, the two integrals from the bremsstrahlung diagram that involve a real final photon are unchanged 
from the electron-photon scattering considered in \cite{KT2018}:
\be
I_2 = - (\varepsilon^z)^* e^{-ik_x(a_I+a_f)/2}e^{-\lambda_B^2k_\perp^2/4}\,
\frac{p_{z,I}(E_{f} + m) + p_{z,f}(E_I+m) }{2L^2 [E_I E_{f} (E_I+m) (E_f+m)]^{1/2}}
\, (2 \pi)^2 \delta\left ( \frac{a_f-a_I}{\lambda_B^2}-k_{y} \right ) \, \delta(p_{z,I} - p_{z,f} - k_{z}),
\ee
and
\be
I_4 =  i(\varepsilon^z)^* e^{-ik_x(a_I+a_f)/2}e^{-\lambda_B^2k_\perp^2/4}\,
\frac{p_{z,I}p_{z,f} + (E_I+m)(E_f+m)}{2L^2 [E_I E_{f} (E_I+m) (E_f+m)]^{1/2}}
\, (2\pi)^2 \delta\left ( \frac{a_f-a_I}{\lambda_B^2}-k_{y} \right ) \, \delta(p_{z,I} + p_{z,f} + k_{z}).
\ee
Here $k_\perp^2 \equiv k_x^2 + k_y^2$.

\section{Erratum}

This erratum is printed separately in the Astrophysical Journal, and is included for
convenience as an Appendix in this arXiv version of the paper.  In summary, 
the cross section for radiative recombination derivedin Section \ref{s:recomb} must be corrected
downward by a factor of 2.

\subsection{Radiative Recombination}

The rate for $e^- + p \rightarrow H + \gamma$, with both electrons and protons confined to the lowest Landau level,
has been obtained using a detailed balance argument.
The left-hand side of Equation (\ref{RecombRateBalance}) must be augmented by a factor of 2 to account for 
incident electrons of both positive and negative longitudinal momentum $p_z$, which are both created in the inverse
process of photoionization.  Then the generalized Milne relation (\ref{eq:milne}) is reduced by a factor of ${1\over 2}$,
\be
\frac{d \sigma_{\rm fb}}{d(\cos\theta)} = \sigma_{\rm bf} \frac{\omega^2}{2e B},
\ee
as are the differential and total cross sections for radiative recombination,
\ba
\frac{d \sigma_{\rm fb}}{d(\cos\theta)} 
&=& 4 \pi {r_e^2\over\alpha_{\rm em}}  
\left(\frac{\omega m}{e B}\right)  \left ( \frac{|E_{\rm H,0}|}{\omega}  \right )^{3/2}
\left( 2 \sin^2\theta +\frac{\omega}{2 \omega_B} \cos^2 \theta \right);\nn
\sigma_{\rm fb} &=& 8 \pi {r_e^2\over\alpha_{\rm em}}\,\left({\omega/m\over B/B_{\rm Q}}\right)
 \left ( \frac{|E_{\rm H,0}|}{\omega}  \right )^{3/2}\left( \frac{4}{3} +\frac{\omega}{6 \omega_B} \right),
\ea
and the recombination Gaunt factor,
\begin {equation}
g_R(T,\omega) = \frac{3\pi \omega^2}{\alpha_{\rm em}^2 e B}\left ( \frac{|E_{\rm H,0}|}{\omega} \right )^{3/2} 
\exp\left[{|E_{\rm H,0}|\over T}\right]
\left(\frac{4}{3} + \frac{\omega}{6 \omega_B} \right)\quad\quad(\omega \geq |E_{\rm H,0}|).
\end {equation}
This reduction in the free-bound cross section reinforces our conclusion that thermal free-bound emission is
subdominant to thermal bremsstrahlung; the revised Figure \ref{RecombGauntCompare} is shown here.
\begin{figure}
\epsscale{0.85}
\plotone{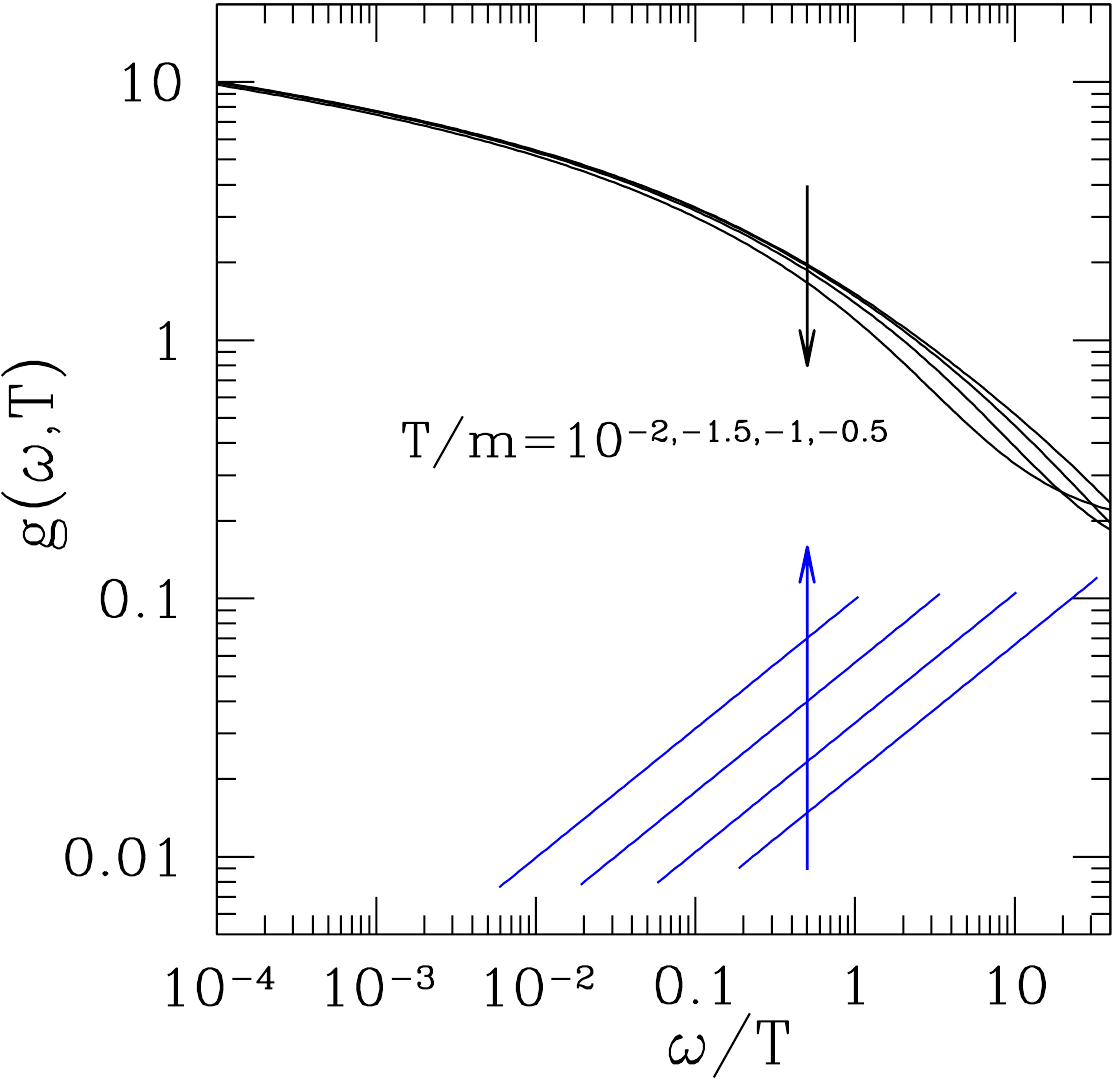}
\caption{Revision to Figure \ref{RecombGauntCompare}.  Comparison of the bremsstrahlung Gaunt factor $g(T,\omega)$
(black curves) and the recombination Gaunt factor $g_R(T,\omega)$ (blue curves) for a range of subrelativistic
temperatures and $B = 100 B_{\rm Q}$.  The arrows point in the direction of increasing $T$.  Free-bound emission
is significantly weaker than free-free emission except at high frequencies, where the emission is exponentially
suppressed.  The recombination curves are reduced by a factor of ${1\over 2}$ compared with those originally published.}
\label{RecombGauntCompare2}
\end{figure}

\subsection{Gaunt Factor Convention}

The normalization of the free-free Gaunt factor given in Equation (\ref{BremssGauntDef}) is chosen to give a simple
analytic form for the Gaunt factor in the nonrelativistic regime (Equations (\ref{LowOmGaunt}) and (\ref{LowTMidOmGaunt})).
It differs from the conventional normalization for nonmagnetized free-free emission.  We also quote the standard
analytic approximations to the nonmagnetized Gaunt factor at the end of Section \ref{ThermBremss} without commenting on this
difference in normalization.  Using our normalization of the Gaunt factor, the nonmagnetic results would
increase by a factor of $2\pi/\sqrt{3}$.  This means that the free-free emissivity of a 
nonrelativistic, but strongly magnetized, thermal plasma is weaker by a factor of ${1\over 2}$ than the standard 
nonmagnetic result.

\subsection{Other Correction}

Following the erratum of \cite{KT2018}, Equation (\ref{eq:31}) must increase by a factor of 2.

\end{document}